\documentclass[epjST]{svjour}
\usepackage{graphicx}
\usepackage{bm}
\usepackage{epsf}
\usepackage{rotating}
\usepackage{epsfig,rotate,color}
\usepackage{wrapfig}
\usepackage{amssymb}
\usepackage{amsmath}
\usepackage{amsfonts}
\usepackage{array,hhline,dcolumn}%
\usepackage{gensymb}
\usepackage{placeins}
\usepackage{pdfpages}
\usepackage[colorlinks=true,citecolor=blue,linkcolor=blue]{hyperref}
\usepackage{accents}
\usepackage{color}
\usepackage{xcolor}
\usepackage[normalem]{ulem}
\usepackage{lineno}
\usepackage{siunitx}
\usepackage{subfig}
\usepackage{multirow, makecell}
\usepackage{booktabs}
\usepackage[rightcaption]{sidecap}
\usepackage[normalem]{ulem}

\newcommand{\nusquids}{\texttt{nuSQuIDS}}
\newcommand{\likelihood}{$\mathcal{L}(\vec\theta; \vec\eta, \vec\eta_b | \vec k)$}
\newcommand{\tgmxs}{1.30^{+0.21}_{-0.19} {\rm(statistical)}^{+0.39}_{-0.43} {\rm (systematic)}}

\def\nue{\nu_e}
\def\numu{\nu_\mu}
\def\nutau{\nu_\tau}
\def\nubar{\bar\nu}
\def\nuebar{\bar\nu_e}
\def\numubar{\bar\nu_\mu}
\def\nutaubar{\bar\nu_\tau}

\def\NDOM{5,160}

\def\NStringI{79}
\def\NStringD{7}
\def\BString{1,450}
\def\EString{2,450}
\def\DHole{125}
\def\NDOMperString{60}
\def\PMTsize{25.4}
\def\DClowE{6} 

\def\Nabs{10,784} 
\def\absangres{0.6^\circ}
\def\absengres{2}
\def\mDOMString{115}
\def\mDOMperString{18}
\def\mDOMNPMT{31}

\def\nunubarRC{0.77}
\def\nunubarRH{+0.44}
\def\nunubarRL{-0.25}

\usepackage{tikz}
\usepackage{environ}

\usetikzlibrary{calc,trees,positioning,arrows,chains,shapes.geometric,%
    decorations.pathreplacing,decorations.pathmorphing,shapes,%
    matrix,shapes.symbols,backgrounds,fit} 
\usepackage{varwidth}

\makeatletter
\newsavebox{\measure@tikzpicture}
\NewEnviron{scaletikzpicturetowidth}[1]{%
  \def\tikz@width{#1}%
  \begin{lrbox}{\measure@tikzpicture}%
  \BODY
  \end{lrbox}%
  \pgfmathparse{#1/\wd\measure@tikzpicture}%
  \BODY
}
\makeatother

\tikzset{
	>=stealth',
	box/.style={
		rectangle,
		rounded corners,
		dashed,
		draw=black, very thick,
		minimum height=2em,
		text centered,
		execute at begin node={\begin{varwidth}{28em}},
		execute at end node={\end{varwidth}}},
	solidbox/.style={
		rectangle,
		rounded corners,
		draw=black, very thick,
		minimum height=2em,
		text centered,
		execute at begin node={\begin{varwidth}{28em}},
			execute at end node={\end{varwidth}}},
    bigsolidbox/.style={
		rectangle,
		rounded corners,
		draw=black, very thick,
		minimum height=6cm,
		text centered,
		execute at begin node={\begin{varwidth}{28em}},
			execute at end node={\end{varwidth}}},
	fw_arrow/.style={
		->,
		thick,
		shorten <=2pt,
		shorten >=2pt,},
	bw_arrow/.style={
		<-,
		thick,
		shorten <=2pt,
		shorten >=2pt,}
}

\definecolor{mc_gen_color}{RGB}{250,138,31}

\definecolor{det_sim_color}{RGB}{227,66,55}

\definecolor{llh_color}{RGB}{128,0,128}

\tikzstyle{bigboxGeneration} = [draw=mc_gen_color!50, thick, fill=mc_gen_color!20, rounded corners, rectangle]

\tikzstyle{bigboxDetector} = [draw=det_sim_color!50, thick, fill=det_sim_color!20, rounded corners, rectangle]

\tikzstyle{bigboxAnalysis} = [draw=llh_color!50, thick, fill=llh_color!20, rounded corners, rectangle]

\begin{document}
%
\title{Neutrino Interaction Physics in Neutrino Telescopes}
\author{Teppei Katori\inst{1}\fnmsep\thanks{
\email{teppei.katori@kcl.ac.uk}}
\and
Juan~Pablo~Yanez\inst{2}\fnmsep\thanks{
\email{j.p.yanez@ualberta.ca}}
\and
Tianlu Yuan\inst{3}\fnmsep\thanks{
\email{tyuan@icecube.wisc.edu}}
}
\institute{
Dept.~of Physics, King's College London, WC2R 2LS London, UK
\and
Dept.~of Physics, University of Alberta, Edmonton, Alberta, T6G 2E1, Canada
\and
Dept.~of Physics and Wisconsin IceCube Particle Astrophysics Center, University of Wisconsin, Madison, WI 53706, USA
}
\abstract{
Neutrino telescopes can observe neutrino interactions starting at GeV energies by sampling a small fraction of the Cherenkov radiation produced by charged secondary particles. These experiments instrument volumes massive enough to collect substantial samples of neutrinos up to the TeV scale as well as small samples at the PeV scale. This unique ability of neutrino telescopes has been exploited to study the properties of neutrino interactions across energies that cannot be accessed with man-made beams. Here we present the methods and results obtained by IceCube, the most mature neutrino telescope in operation, and offer a glimpse of what the future holds in this field.
} 
\maketitle

\section{Introduction}

Neutrino telescopes are large-volume, massive, sparsely instrumented detectors that monitor a natural medium for signatures of neutrino interactions. These interactions are detected via the Cherenkov radiation emitted by the charged secondaries produced in them; the instrumentation samples a small fraction of the emitted radiation and its time of arrival and intensity at each sensor is used to infer the particles present and their properties.

At low energies, the sparseness of the sensor array dictates its energy threshold. At high energies, the sensitivity is limited by low neutrino fluxes and the challenge of instrumenting ever larger volumes. Beyond PeV energies, UV Cherenkov detectors are no longer practical, and other technologies such as radio or acoustic detection are necessary~\cite{Katz:2011ke}.

For most of the energy range that neutrino telescopes cover -- a few GeV and above -- neutrinos interact mainly with nucleons via deep inelastic scattering (DIS). One exception happens near the detector threshold, where neutrinos can be detected after quasi-elastic (QE) interactions with nucleons or after exciting a nucleon to a resonant state. Another special case occurs in a narrow energy range around \SI{6.3}{\peta \eV}, where $\nuebar - e^-$ interactions dominate to produce an on-shell $W^-$~boson~\cite{Glashow:1960zz}.

Accelerator-based cross section measurements exist up to a few hundreds of GeV~\cite{Formaggio:2013kya}. In this energy region, uncertainties in cross sections impact the understanding of oscillation data and need to be accounted for. Above $\sim$\SI{300}{\giga \eV}, however, neutrino telescopes serve as our only probe of weak interactions. Moreover, due to the large detection region of these experiments, they can also be used to search for rare weak processes at all energies, such as neutrino trident production~\cite{Ge:2017poy}.

The largest neutrino telescope in operation to date, the IceCube Neutrino Observatory~\cite{Aartsen:2016nxy}, uses ice as its Cherenkov medium. While this review focuses on IceCube results, the discussions are applicable to other projects. In Sec.~\ref{sec:icecube_data} we begin by describing the basics of the IceCube detector, expected signals and event reconstruction, and the general analysis strategy that these studies follow. In Sec.~\ref{sec:nusim} we cover how neutrino cross sections are modeled, while Sec.~\ref{sec:nuflux} goes over the origin of the neutrinos fluxes used. Section~\ref{sec:results} describes cross section measurements and studies where their uncertainties play a major role. We conclude in Sec.~\ref{sec:future} with a discussion of future projects and their relevance for neutrino interaction studies.

\section{The data of the IceCube Neutrino Observatory}
\label{sec:icecube_data}

\subsection{The IceCube Neutrino Observatory}

\begin{figure}[hbt]
\centering
\includegraphics[width=0.54\columnwidth]{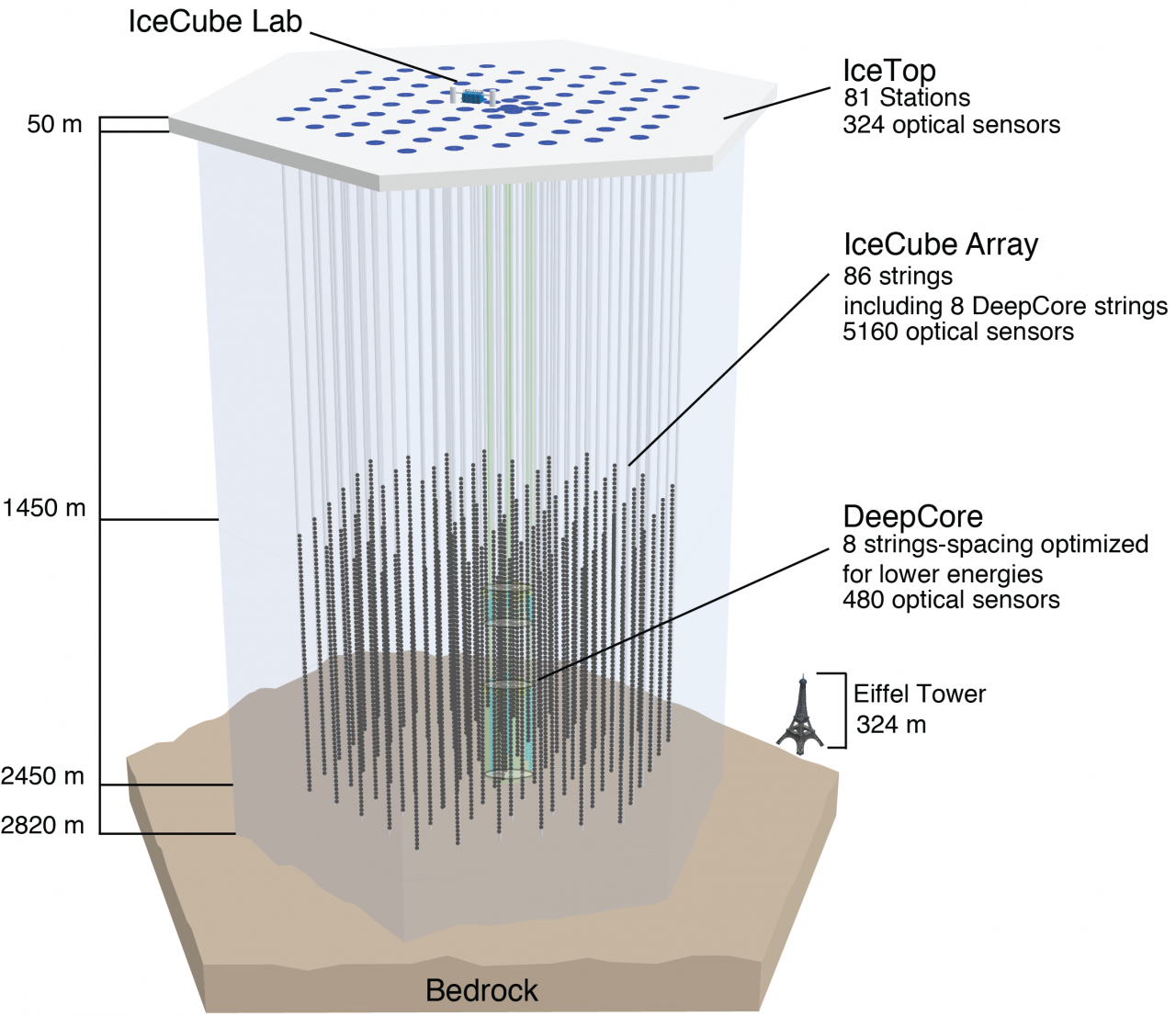}
\includegraphics[width=0.45\columnwidth]{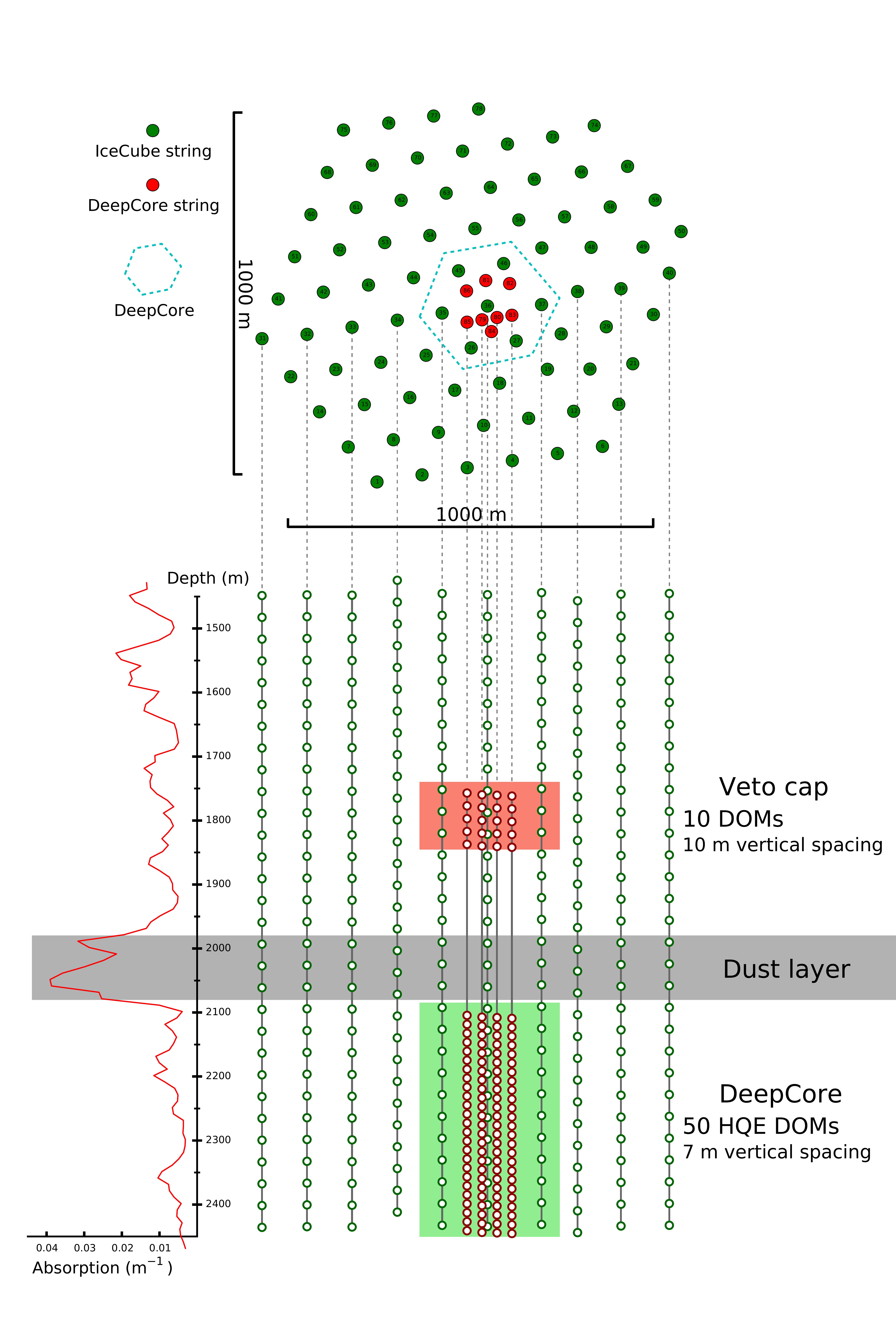}
\caption{Left: The IceCube Neutrino Observatory~\cite{Aartsen:2016nxy}. The IceTop surface array is shown with the IceCube in-ice array, including DeepCore. Right: Top and side views of the detector~\cite{Aartsen:2019tjl}. The red curve on the left shows optical absoprtion coefficient as a function of depth. A dust layer (grey band) splits the detector in two.}
\label{fig:detector}
\end{figure}

The IceCube Neutrino Observatory~\cite{Aartsen:2016nxy}, illustrated in Fig.~\ref{fig:detector}, is located at the geographic South Pole, Antarctica. The glacial ice and the bedrock below serve as interaction targets for incoming neutrinos. A downward-facing $\PMTsize$~cm photo-multiplier tube (PMT) and accompanying electronics are enclosed in a pressure resistant glass sphere to make a digital optical module (DOM), the detection unit of IceCube.

The $\NDOM$ DOMs that form the array are distributed within roughly 1~km$^3$ of ice. Here, $\NDOMperString$ DOMs are attached on each string, which were lowered into a hole drilled in the ice during deployment. DOMs are located at depths between $\BString$ to $\EString$~m from the top of the ice sheet, and each string is separated roughly by $\DHole$~m to make a hexagonal grid of $\NStringI$ strings. $\NStringD$ strings are distributed near the center, with closer string separation to detect lower-energy particles, in a volume called DeepCore~\cite{Collaboration:2011ym}. Analyses are typically classified as \textit{low energy} ($\leq 100$~GeV) if they mainly rely on DeepCore data. 

The large difference in sensor spacing between vertical and horizontal planes can be seen in Fig.~\ref{fig:detector}, right, which shows the top and the side views of the detector~\cite{Aartsen:2019tjl}. This, together with PMTs that point downwards, makes the detector performance depend significantly on the position and direction of the particles observed. Figure~\ref{fig:detector} also shows a layer of dust with very high scattering and absorption that lies at a depth between approximately \SIrange{1980}{2080}{\m}~\cite{paleowind,IceCube:2013llx}. Analyses will typically avoid this region due to its poor optical properties.

\subsection{Interaction kinematics and event reconstructions}
\label{sec:reco}

\begin{figure}[!b]
\centering
\includegraphics[width=0.95\columnwidth]{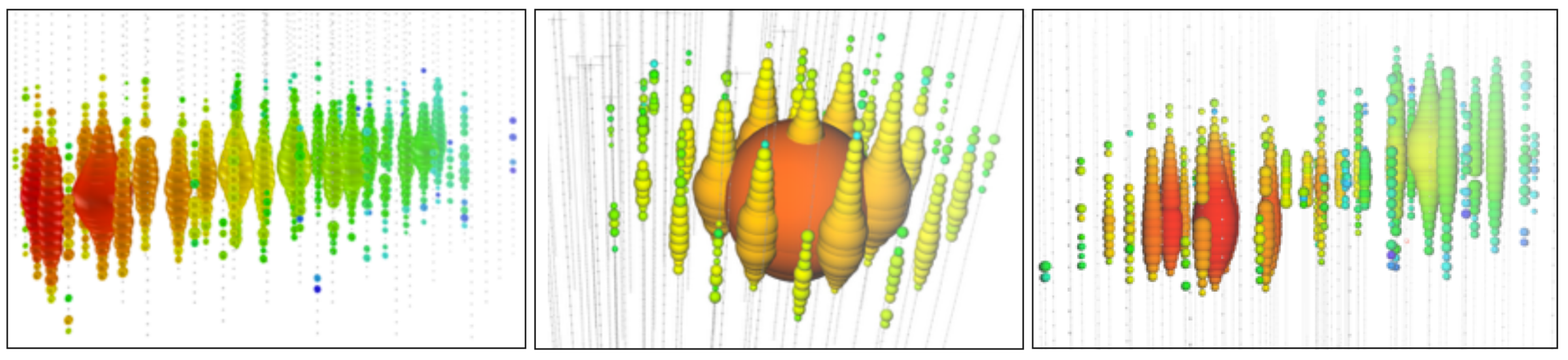}
\includegraphics[width=0.6\columnwidth]{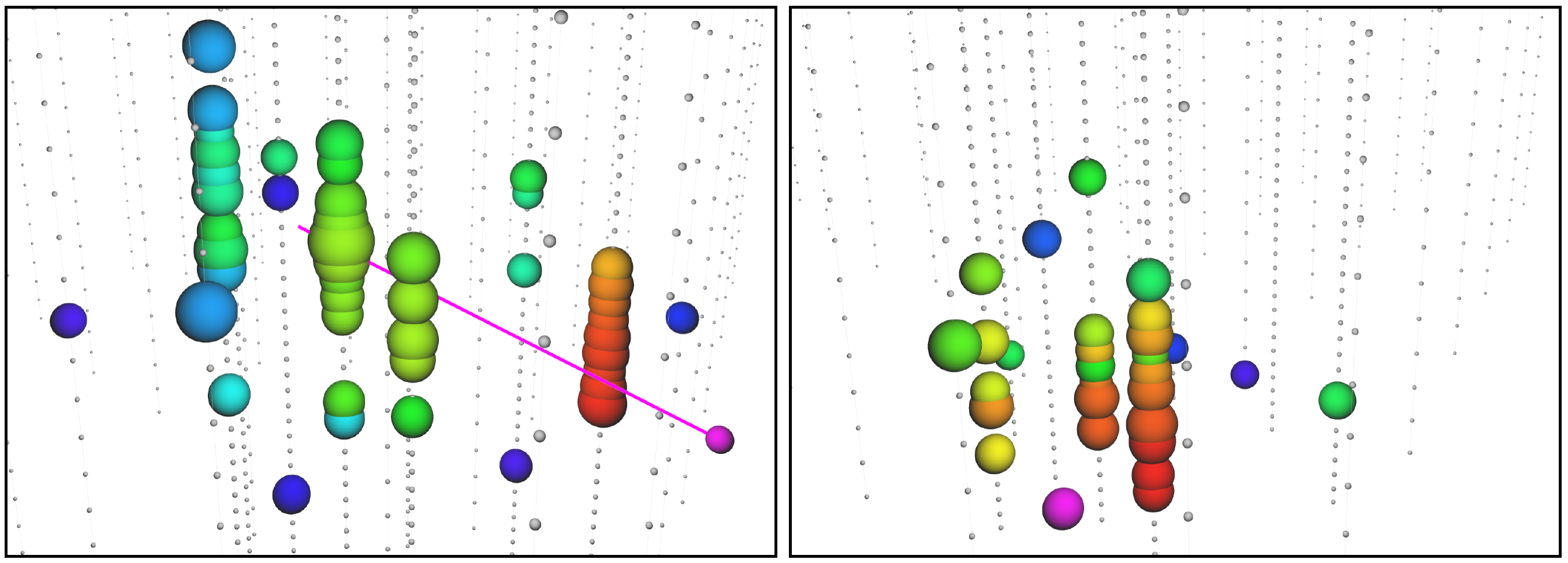}
\caption{Examples of neutrino interaction signatures in IceCube. DOMs that have observed light are represented as spheres with their size proportional to the number of photo-electrons detected and their color representing the time of the first hit going from red (early) to blue (late). IceCube events are shown on the top panels~\cite{Aartsen:2020fgd} while bottom panels correspond to low energy events in DeepCore~\cite{Terliuk:2018xom,summer_blot_2020_4156203}.
IceCube event displays include ``track'' (top left), ``cascade'' (top middle), and simulated ``double-cascade'' (top right). Low energy event displays are for simulated ``track'' (bottom left) and ``cascade'' (bottom right).}
\label{fig:event}
\end{figure}

The light collected by the DOMs is used to establish the type of charged particle identified (PID), its energy and its direction~\cite{IceCube:2013dkx}. After that, different signal hypotheses might be used to determine the properties of the neutrino responsible for the particles observed. Examples of neutrino-interaction signatures in IceCube are shown in Fig.~\ref{fig:event}. Note, neutrinos and antineutrinos are indistinguishable by neutrino telescopes with a few exceptions.

Muons above tens of GeVs travel distances long enough to produce a noticeable ``track-like'' pattern (Fig.~\ref{fig:event}, top and bottom left) in a sparse detector like IceCube, so this topology is used to distinguish them from other particles. Their extension together with the arrival time of their Cherenkov light are crucial for establishing their travel direction. Their energy is estimated from a combination of track length and brightness, and the method depends on the energy scale of the muon and whether the track is contained in the instrumented volume. 

Muons with an energy below 700~GeV in ice lose their energy mainly via ionization~\cite{Kowalski:2004qc}, a constant process that allows the muon to travel roughly 4.5~m/GeV. Above this energy, stochastic radiative processes that scale with the muon energy dominate~\cite{IceCube:2021oqo}. The energy of contained muons is thus estimated mainly by their range, which can be practically done up to 200~GeV. Once the muons are not contained energy estimates rely on their stopping power. The energy reconstruction from stochastic energy deposit can provide a lower limit on the through going muon energy~\cite{IceCube:2013dkx} and the resolution can be refined by detailed treatments of dE/dx~\cite{IceCube:2012iea}.

The directional resolution for muon tracks goes from 5$^\circ$ to 10$^\circ$ near the low energy threshold~\cite{Aartsen:2014yll} to sub-1$^\circ$ above 1~TeV~\cite{IceCube:2021oqo}. The energy estimation for contained muons in the minimum ionizing regime has an error of about 10\,GeV~\cite{YanezGarza:2014jia}, 
while in the radiative regime muon energy estimators range from 40\% error at 3~TeV to about 30\% at 1~PeV~\cite{IceCube:2013dkx}.

Apart from muons, all other charged leptons and hadrons initiate a cascade of secondaries. These cascades develop over a few meters in the longitudinal direction, and the secondaries spread in the transverse direction, so for a neutrino telescope they appear point-like (Fig.~\ref{fig:event}, top middle and bottom right). This results in a significant spread in the overall light direction making their directional reconstruction more challenging than for tracks, giving a directional resolution of approximately 15$^\circ$, depending on the energy range~\cite{IceCube:2013dkx,Aartsen:2017nmd}. On the other hand, due to their localization, the deposited energy resolution improves with energy, starting with 30\% below 100~GeV~\cite{Aartsen:2017nmd} down to 8\% at 100~TeV~\cite{IceCube:2013dkx}.

The flavor and kinematic properties of the neutrino responsible for an observed event are reconstructed by putting together the pieces described above. The presence of a track indicates a $\numu$($\numubar$) charged-current (CC) interaction, while all other DIS interactions result in cascades. Muon neutrinos with an interaction vertex inside the detector thus contain the most information. For these interactions the neutrino energy can be estimated by summing up muon and hadronic shower energies ($E_\nu=E_l+E_h$), the invariant mass of the hadronic system is $W=\sqrt{2E_\nu M}$, where the target is usually assuming nucleons at rest with mass $M$, the inelasticity $y$ is $y=E_h/E_\nu$ and their direction can be estimated by that of the muon. 

All neutral-current (NC) interactions, $\nue(\nuebar)$CC, and most $\nutau$($\nutaubar$)CC cannot be distinguished, and the neutrino energy can only be estimated after assuming a particle hypothesis from an expected flux. Interactions of $\nutau$($\nutaubar$)CC can be an exception when the high-energy $\tau$ lepton carries enough energy to travel a few meters ($\sim50$~m$\cdot E_\tau$/PeV) before decaying, producing a double cascade signature (Fig.~\ref{fig:event}, top right). Two candidates have been identified already using this method~\cite{Aartsen:2015dlt,Usner:2018cel,Abbasi:2020zmr}. Searching for light ``echo'' has also been proposed as a statistical method for identifying $\nutau(\nutaubar)$CC~\cite{Li:2016kra,Steuer:2018qmb}. 
Similarly, distinction between particles and antiparticles is not possible on an event-by-event basis, and can only be done statistically, although there are on-going efforts to tag small subsets of events using decay times of muonic atoms~\cite{PhysRev.79.749,singhal1983determination,Knecht:2020npz}.

\subsection{Data analysis strategy}

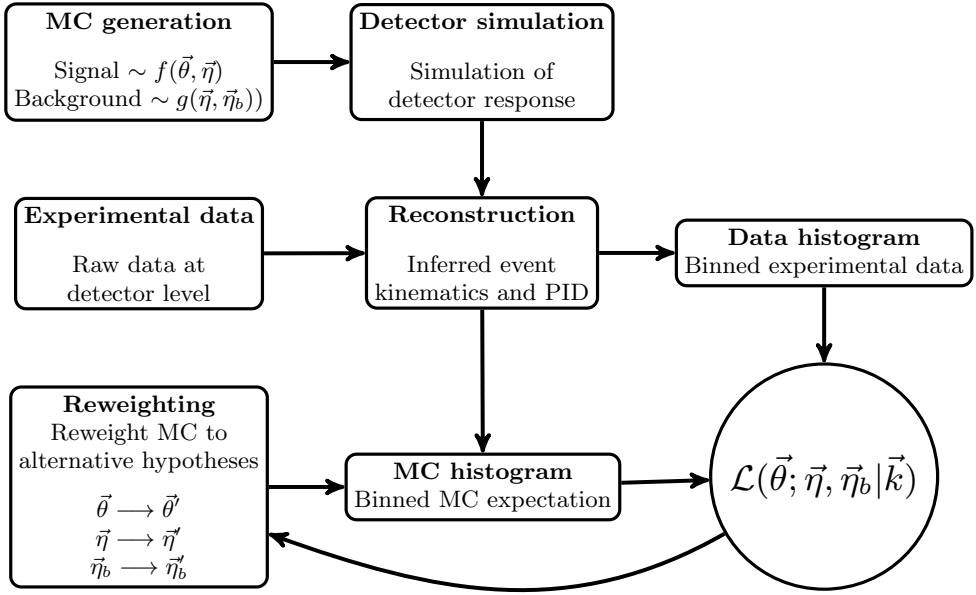
\begin{figure}[htp]                                                                                                    
\resizebox{\textwidth}{!}{
\begin{tikzpicture}[node distance=1cm, auto,]

\node[solidbox] (mc_generation) {\centering {\bf MC generation} \\
	\begin{center}
		Signal $\sim f(\vec\theta, \vec \eta)$ \\
		Background  $\sim g(\vec\eta, \vec\eta_b)$)
	\end{center}
};

\node[solidbox, below=of mc_generation] (data) {\centering {\bf Experimental data}\\
	\begin{center}
        Raw data at\\detector level
	\end{center}
};

\node[solidbox, right=of mc_generation] (detector_simulation) {\centering {\bf Detector simulation} \\
    \begin{center}
	 Simulation of\\detector response
	\end{center}
};

\node[solidbox, below=of detector_simulation] (reconstruction) {\centering {\bf Reconstruction} \\
    \begin{center}
	 Inferred event\\kinematics and PID
	\end{center}
};

\node[solidbox,below=of data] (reweighting) {\centering {\bf Reweighting} \\
	Reweight MC to\\alternative hypotheses \\
	
	\begin{center}
    	$\vec\theta \longrightarrow \vec\theta'$\\
    	$\vec\eta \longrightarrow \vec\eta'$\\
    	$\vec\eta_b \longrightarrow \vec\eta_b'$
	\end{center}
};

\node[solidbox,right=of reweighting] (mc_histogram) {\centering {\bf MC histogram} \\
	Binned MC expectation
};

\node[solidbox,right=of reconstruction] (data_histogram) {\centering {\bf Data histogram} \\
	Binned experimental data
};

\node[circle, minimum size=3cm, color=black, draw=black, very thick, below=of data_histogram] (likelihood) {\centering {\scalebox{1.5}{\likelihood}}
};


\draw [->,line width=1.5pt] (mc_generation) -- (detector_simulation);
\draw [->,line width=1.5pt] (detector_simulation) -- (reconstruction);
\draw [->,line width=1.5pt] (data) -- (reconstruction);
\draw [->,line width=1.5pt] (reconstruction) -- (mc_histogram);
\draw [->,line width=1.5pt] (reconstruction) |- (data_histogram);
\draw [->,line width=1.5pt] (reweighting) -- (mc_histogram);
\draw [->,line width=1.5pt] (mc_histogram) -- (likelihood);

\draw [->,line width=1.5pt] (data_histogram) -- (likelihood);
\draw [->,line width=1.5pt] (likelihood) to [out=-150,in=-20] (reweighting);

\end{tikzpicture}
}
\caption{Forward-folding flowchart as adapted from~\cite{Arguelles:2019izp}. In order to measure signal parameters, $\vec\theta$, a maximum-likelihood approach is used based on MC simulation and event-by-event reweighting. At the generation step, physical processes are simulated for signal (background) according to some distribution $f(\vec\theta, \vec\eta)$ ($g(\vec\eta, \vec\eta_b)$), where $\vec\eta$ ($\vec\eta_b$) are joint (background-only) nuisance parameters. The generation output -- typically true particle kinematics and PID -- is then fed into a realistic detector simulation, which yields a representation of experimental, raw data. The representative and experimental data can both be reconstructed to obtain inferred event kinematics and PID, which are histogrammed in order to construct a likelihood, \likelihood. The MC can be reweighted to different physical hypotheses as signal and nuisance parameters are varied.}
\label{fig:mc_diagram}                                                                                                 
\end{figure}

IceCube analyses require Monte Carlo (MC) simulation sets that are compared with data. This process begins by producing enough statistics of each of the types of interactions expected in the detector and, in some cases, sets that mimic possible overlaps. The simulation is subject to processes that mimic the detector response to light, triggering and filtering. After that, data and simulation are treated in the exact same way, going through the same event selection and reconstruction steps. At the end of the process both data and simulation have been affected by any biases introduced by these steps.

A quantitative comparison of simulation to data is achieved by histogramming both sets in the space of these reconstructed quantities, typically track and cascade energy as well as direction, and constructing a likelihood. The histogram of the data in these variables is compared to the one produced from the sum of all the components from the simulation. In cases where there is low statistics in simulation relative to data, modifications to the standard Poisson likelihood can be used to mitigate biases due to MC fluctuations~\cite{Chirkin:2013lya,Arguelles:2019izp}. The simulation is adjusted to match the data using event-by-event weights to find the set of values that result in the best possible match using a maximum likelihood estimator (MLE) -- or maximum a posteriori probability (MAP) estimator in the Bayesian approach. Uncertainties are obtained by exploring the results in the vicinity of the MLE or MAP. The procedure outlined above is sometimes referred to as ``forward folding'' and it is the one used in the studies described here. The full process is depicted in Fig.~\ref{fig:mc_diagram}.

The simulation weights can be modified to introduce effects expected to impact the data, which might come from the physical theory being tested or from imprecise knowledge of the ingredients required to build the simulation. Of particular interest for neutrino telescopes are uncertainties from the detection process, namely the optical properties of the detection medium and the detector response. These detection uncertainties are included by simulating equivalent sets with modified descriptions of detector and medium. The sets are combined to produced parameterizations of these uncertainties~\cite{Aartsen:2019tjl,Aartsen:2019jcj}, which results in detection systematic uncertainties becoming another weighting factor for the simulation.

\section{Neutrino interaction simulation for Neutrino Telescopes}
\label{sec:nusim}

\subsection{Low-energy neutrino interactions}
\label{sec:lowe_xs}

The neutrino interactions observed by DeepCore, the low energy extension, can be as low as $\DClowE$ GeV. Thus, the physics and tools must overlap with neutrino interaction physics in current and future accelerator-based neutrino experiments~\cite{Alvarez-Ruso:2017oui}. To simulate low energy neutrino interactions, GENIE 2.8.6~\cite{Andreopoulos:2009rq} was used for recent DeepCore analyses~\cite{Aartsen:2019tjl,Aartsen:2017nmd}. Here quasi-elastic interactions are based on the Fermi gas model, and the Rein-Sehgal model is used to simulate baryonic resonance (RES) and coherent meson productions including Delta and higher resonances~\cite{Rein:1980wg,Rein:1982pf}. The DIS model uses LO GRV98 PDF~\cite{Gluck:1998xa} with Bodek-Yang correction at the low $Q^2$ region motivated from quark-hadron duality~\cite{Bodek:2002ps,Bodek:2003wc,Bodek:2004pc}. At low $W$ region hadronization model is based on the empirical fits and KNO-scaling~\cite{Koba:1972ng} with neutrino bubble chamber data, but at high $W$, PYTHIA6~\cite{Sjostrand:2006za} is used. The AGKY model~\cite{Yang:2009zx,GENIE:2021wox} is used to connect models in the shallow-inelastic scattering (SIS) region~\cite{Andreopoulos:2019gvw,SajjadAthar:2020nvy} smoothly, from the baryonic resonance to DIS, and low $W$ to high $W$ hadronization models. The physics in this region, in particular QE and RES dominant energy region, is an active research field and many systematic errors are developed by others~\cite{Wilkinson:2016wmz,Acero:2020eit,GENIE:2021zuu}, but for the DeepCore analysis DIS is the dominant channel. Neutrino SIS-DIS events show large data-model discrepancy~\cite{Mousseau:2016snl} in nuclear target and this is not understood. There is a speculation that the target dependence could be large and nuclear dependent PDF may be different in charged leptons and neutrinos~\cite{Hirai:2007sx,Cacciari:2009dp,Schienbein:2009kk,deFlorian:2011fp}. Around 1 to 10 GeV, several cross-section channels contribute to neutrino interactions simultaneously, and it is challenging to connect all models consistently in the simulation.

\subsection{High-energy neutrino interactions}

IceCube relies on the CSMS neutrino interaction model~\cite{CooperSarkar:2011pa} for simulation of events above 100~GeV. In the CSMS model, the cross-section is defined for the isoscalar target and NLO PDFs from CT10~\cite{Lai:2010vv} and HERAPDF1.5~\cite{CooperSarkar:2010wm} are used. Here, sea quarks include up to $b$ quarks. The total cross-section prediction covers from $50$~GeV to $10^{11}$~GeV for both neutrino and antineutrino, CC and NC, and the predicted error of the cross-section is around 2\% at 60~TeV to 10~PeV. In the simulation, the total CSMS cross section is chosen and then it samples the energy and direction of final state daughter leptons according to the standard formulation. A new NLO (CSMS) and NNLO (BGR18~\cite{Gauld:2016kpd,Bertone:2018dse}) framework to extend the GENIE interaction generator to this energy regime is being built ~\cite{Garcia:2020jwr}.

Neutrino telescopes use natural materials as a target. Examples are sea water, the Antarctic ice, and the bedrock, which include heavy nuclear targets and not isoscalar. Therefore the nuclear effect of high-energy neutrinos needs to be evaluated~\cite{Klein:2020nuk}. For this, the structure functions are written for non-isoscalar targets, and the nuclear PDF EPPS16, developed by $p-Pb$ collision data from the LHC~\cite{Eskola:2016oht}, is used. There are many ways these effects can affect cross-section measurements in IceCube. First, neutrinos passing through the Earth core are affected since the Earth core contains heavy elements such as iron. Second, the main target of cross-section measurements are water molecules which are not isoscalar. At $E_\nu\leq 100$~TeV, the non-isoscalar effect is important, but for the higher energy, nuclear effects such as shadowing and antishadowing are more important and they are predicted to be as big as $\sim$4\% for the total neutrino and antineutrino cross-sections~\cite{Klein:2020nuk}. Although these are important effects, they are smaller errors compared with flux systematic errors, discussed next. 

\section{Neutrino flux}
\label{sec:nuflux}

Neutrino telescopes observe neutrinos produced in Earth's atmosphere and, at the highest energies, in astrophysical sources. All results presented in this article include systematic errors of atmospheric or astrophysical neutrino flux. Below 50~TeV the Earth is essentially transparent to neutrinos and the flux is expected to come from all directions, while above it neutrinos traveling through a sufficient chord length in the Earth may interact prior to arriving at the detector.

\subsection{Atmospheric flux of muons and neutrinos}

The vast majority of events that neutrino telescopes record arise from cosmic ray showers in the atmosphere. As these showers develop, the hadronic component produces atmospheric neutrinos and cosmic muons\footnote{For a comprehensive overview of this topic, see \cite{Gaisser:2002jj}}. Cosmic muons are the main source of background for most neutrino studies in neutrino telescopes, triggering these experiments at a rate $10^6$ times higher than neutrinos~\cite{Collaboration:2011ym}. They can be removed by means of their reconstructed direction or by requiring that the reconstructed neutrino vertex is inside the detector volume with no correlated signal in an outer veto region. Most analyses require simulating large cosmic muon datasets to properly include their impact on the signal of interest.

\begin{figure*}[hbt]
  \centering\includegraphics[width=0.7\textwidth]{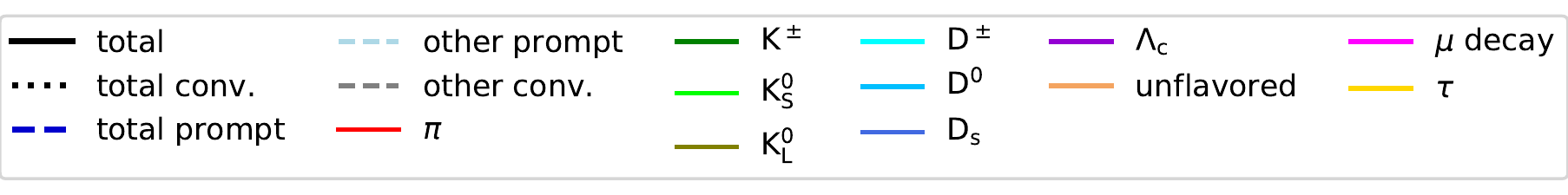}
  \includegraphics[clip, trim=0 0 11.23cm 0., width=0.45\textwidth]{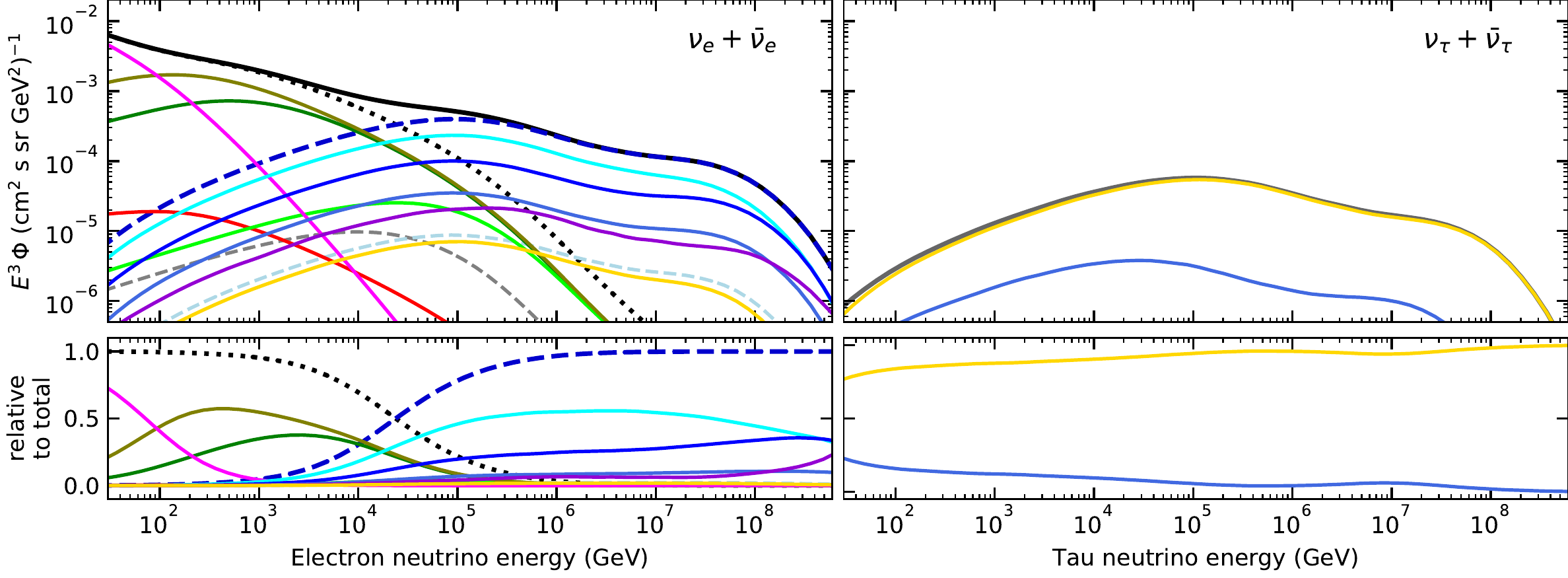}
  \includegraphics[clip, trim=0 0.24cm 22.62cm 0, width=0.054\textwidth]{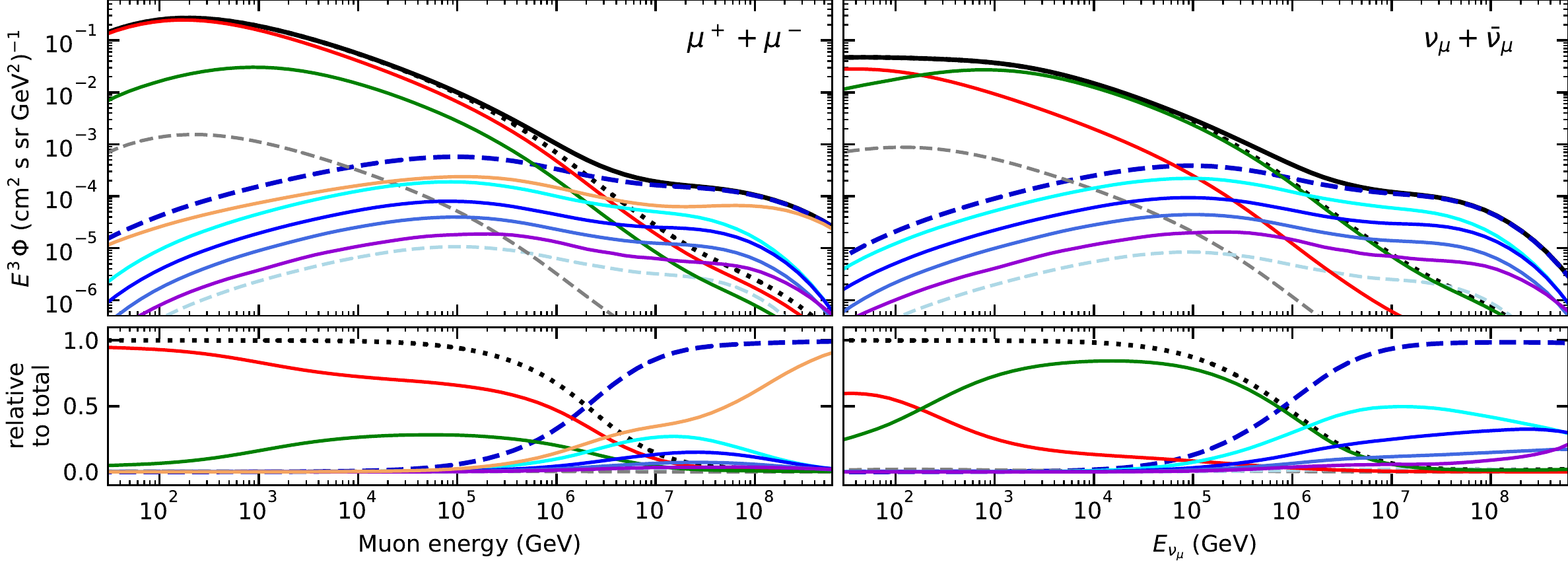}
  \includegraphics[clip, trim=12.95cm 0.24cm 0 0, width=0.4\textwidth]{figures/hadr_breakdown_1.pdf}
  \caption{\label{fig:hadron_contribution} Atmospheric neutrino fluxes for $\nue+\nuebar$ (left) and $\numu+\numubar$ (right), broken down by their parent particle as given by the hadronic interaction model Sibyll-2.3c and the H3a cosmic ray model \cite{Gaisser:2012zz} at $\theta =60^\circ$. Figure reproduced from~\cite{Fedynitch:2018cbl}.}
\end{figure*}

The dominant signal for most studies discussed in this paper are atmospheric electron and muon neutrinos. Just like the cosmic muons, these neutrinos originate from the decay of mesons produced in atmospheric showers. Expected fluxes from calculations are shown in Fig.~\ref{fig:hadron_contribution}, where one can see that muon neutrinos are the dominant component across all energies.

Figure~\ref{fig:hadron_contribution} also contains information on the parent particle for each flavor. The decay of light mesons, $\pi^\pm$, $K^\pm$, $K^0_L$, and $K^0_S$~\cite{Gaisser:2014pda} gives rise to the low energy component ($\leq 50$~TeV) of the spectrum, typically labeled as ``conventional''. These mesons have a long enough half-life to interact with the atmosphere before decaying. This component has been measured up to a few hundred TeV~\cite{Aartsen:2014qna}. The high energy end of the spectrum is expected to be dominated by the decay of $D$ (charmed) mesons, which decay without reinteraction in the atmosphere, and thus is commonly referred to as ``prompt''. This transition occurs between 10~TeV and 1~PeV, depending on the flavor and observation direction. 

Due to its short lifetime, the energy spectrum of the prompt component is harder than the conventional one, as seen in Fig.~\ref{fig:hadron_contribution}. This poses a challenge when separating it from astrophysical neutrinos, described next. Because of this complication, the prompt flux component of atmospheric neutrinos has not yet been conclusively measured.

\subsection{Astrophysical neutrino flux}

Neutrino interactions at energies beyond atmospheric neutrinos ($\gtrsim 60$~TeV) are measured using the astrophysical neutrino flux. These are mainly diffuse, namely the flux is roughly isotropic. For cross section measurements, the flavor structure is typically assumed to be $(\nue:\numu:\nutau)\sim (1/3:1/3:1/3)$ at Earth, consistent with data~\cite{Abbasi:2020zmr,Aartsen:2015ivb,Aartsen:2015knd,Aartsen:2018vez}. There are several candidate processes for the astrophysical neutrino production model, and the preferred production model is from pion decays where pions are produced in either $p-p$ or $p-\gamma$ interactions. Similar to atmospheric neutrinos, this scenario predicts an initial flavor ratio of $(\nue:\numu:\nutau)\sim (1/3:2/3:0)$. However, due to the long propagation distance ($>> 1Mpc$), large neutrino production source ($>> L_{osc}$)), and poor energy resolution ($\Delta E/E>> 1\%$)
the arrival flux at Earth is incoherent and flavor-mixed~\cite{Farzan:2008eg}. From the large mixing angle between $\numu$ and $\nutau$, we expect roughly equal contributions for all flavors. This situation is rather insensitive to the assumed astrophysical neutrino production model~\cite{Arguelles:2015dca,Bustamante:2015waa}.

The spectrum of astrophysical neutrinos is still being understood~\cite{Aartsen:2018vez,Abbasi:2020jmh}. A standard assumption is that it is a single power law spectrum for all flavors, $\sim\Phi_{astro}\cdot E^{-\gamma_{astro}}$ where $\Phi_{astro}$ is the flux normalization and $\gamma_{astro}$ is the spectral index, which is measured between 2 and 3, depending on the sample~\cite{Aartsen:2018vez,Abbasi:2020jmh}. To accommodate this large uncertainty, it is common to use  $\Phi_{astro}$ and $\gamma_{astro}$ as nuisance parameters in analyses, and the high-energy neutrino cross-sections are measured with simultaneous constraint of the astrophysical neutrino flux model.

\subsection{Neutrino scattering in the Earth}

\begin{figure}[h!]
\centering
\includegraphics[width=0.90\columnwidth]{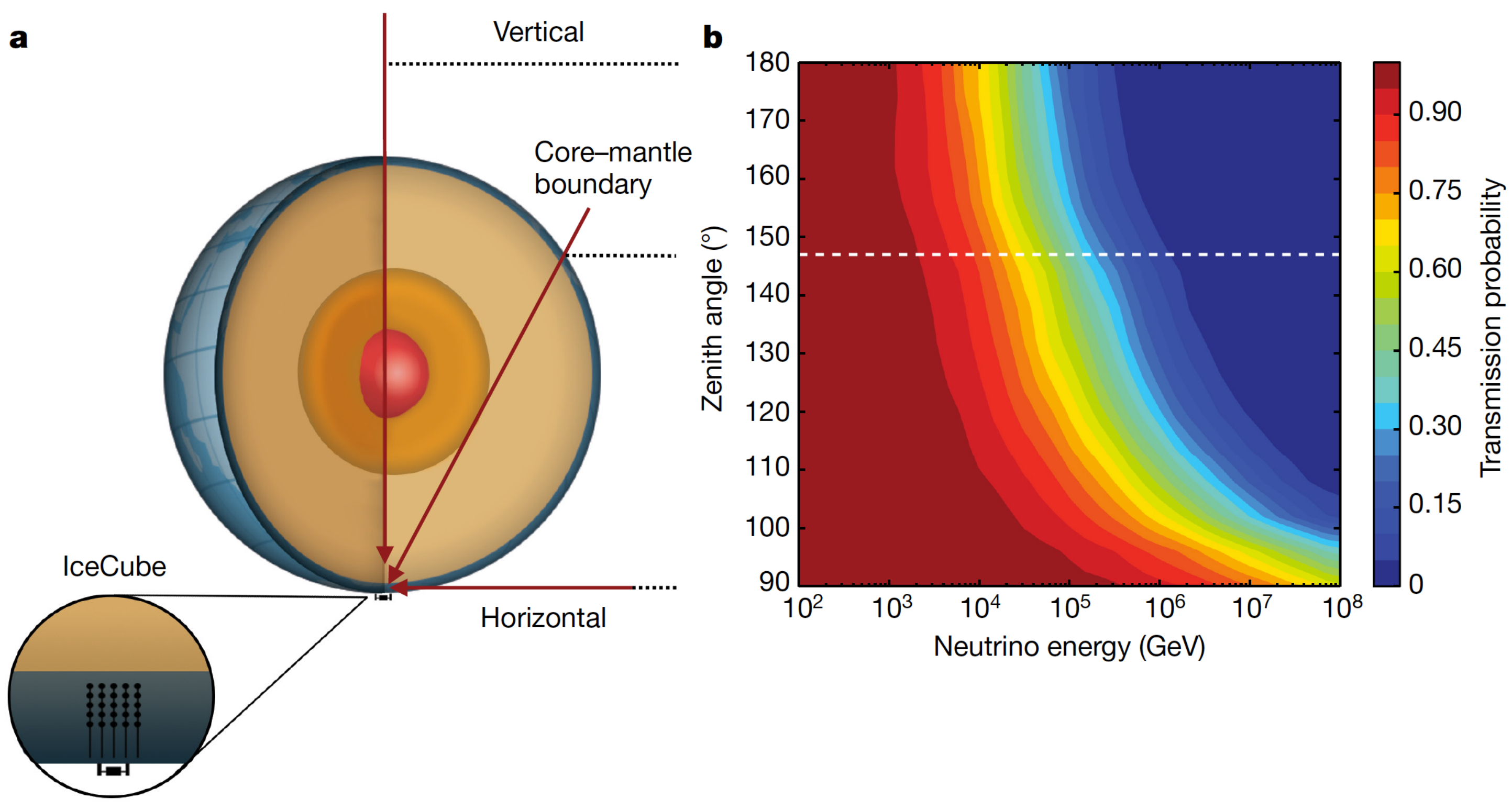}
\caption{Schematic from Ref.~\cite{Aartsen:2017kpd} showing the implications of neutrino interactions in the Earth on the expected IceCube arrival flux. The neutrino-earth material scattering makes the zenith-dependent effect. This becomes important at energies above approximately \SI{1}{\tera \eV}.}
\label{fig:absorption}
\end{figure}

Since the DIS cross-section increases with energy, as energy increases the Earth is no longer transparent for neutrinos. Thus, by assuming a neutrino flux and a number of target nucleons along the neutrino trajectory, the measured neutrino rate can be used to find the total neutrino-nucleon cross section. To do this, it is necessary to assume the density profile of the Earth. The PREM (Preliminary reference earth model)~\cite{Dziewonski:1981xy} is typically used for this. The PREM is widely used in long-baseline neutrino oscillation experiments to take into account the matter effect in neutrino oscillations~\cite{Aartsen:2017nmd,Abe:2019vii,Acero:2019ksn,Jiang:2019xwn} and high-energy neutrino data are consistent with it~\cite{Donini:2018tsg}. Neutrinos producing up-going muons in the IceCube coordinate system are generated at the northern hemisphere (Fig.~\ref{fig:absorption}, left), and their flux is attenuated at the high energy. At 50~TeV, roughly 50\% of neutrinos from the North Pole (vertically up-going neutrinos in IceCube coordinates) are scattered and do not pass through the Earth (Fig.~\ref{fig:absorption}, right). 

\section{Neutrino interaction measurements with IceCube data}
\label{sec:results}

The diffuse fluxes that IceCube measures have been used for obtaining results both at the low and high energy regimes. At low energies, where direct measurements exist, variations of the neutrino-nucleon cross section do not have a large impact on oscillation results and are instead considered a source of systematic uncertainty. This might change with future analyses and as more data on $\nutau$ events are collected. At high energies, with no other measurements available, IceCube data can be uniquely used to probe the cross section.

\subsection{Low energy neutrino analysis}

IceCube low energy analyses rely on data from the DeepCore sub-array, where events with energies down to 6\,GeV can be detected. Low energy studies using atmospheric neutrinos as signal have mostly focused on measuring the $\numu+\numubar$ spectrum below 100\,GeV. In this regime the signature of standard oscillations is expected to be remarkably strong, with (almost) complete disappearance of $\numu(\numubar)$ at 25\,GeV for neutrino trajectories that cross the entire Earth. The strong disappearance, together with the large statistics and the ability to map a very large space in $L/E$ over which oscillations develop has made it possible for IceCube to report precise measurements of the mixing angle and mass splitting that govern atmospheric oscillations, namely $\theta_{23}$ and $\Delta m^2_{32}$~\cite{Aartsen:2014yll,Aartsen:2017nmd,Aartsen:2019tjl}. 

\begin{figure}
    \centering
    \begin{minipage}[c]{0.53\textwidth}
        \centering
        \includegraphics[clip, trim=0.8cm 0 1cm 0.1cm, width=\linewidth]{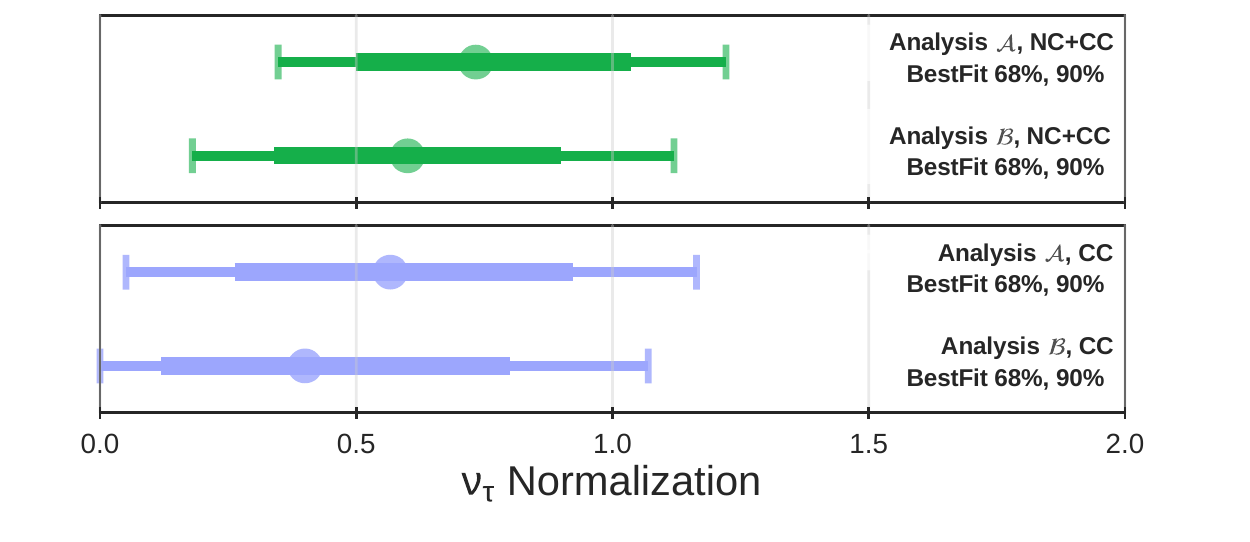}
        \includegraphics[width=\linewidth]{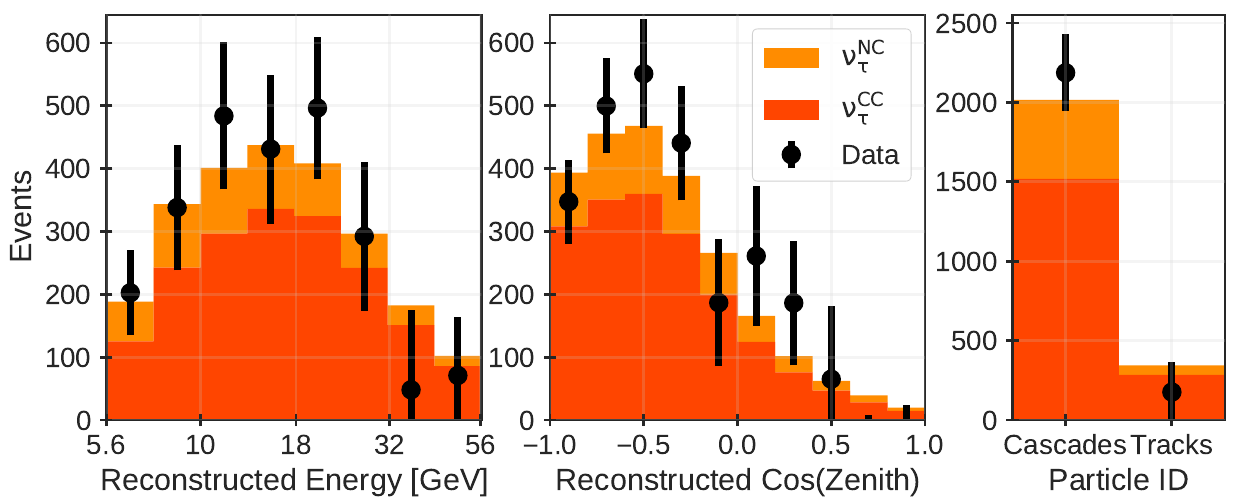}
    \end{minipage}
    \centering
    \begin{minipage}[c]{0.45\textwidth}
        \includegraphics[width=\linewidth]{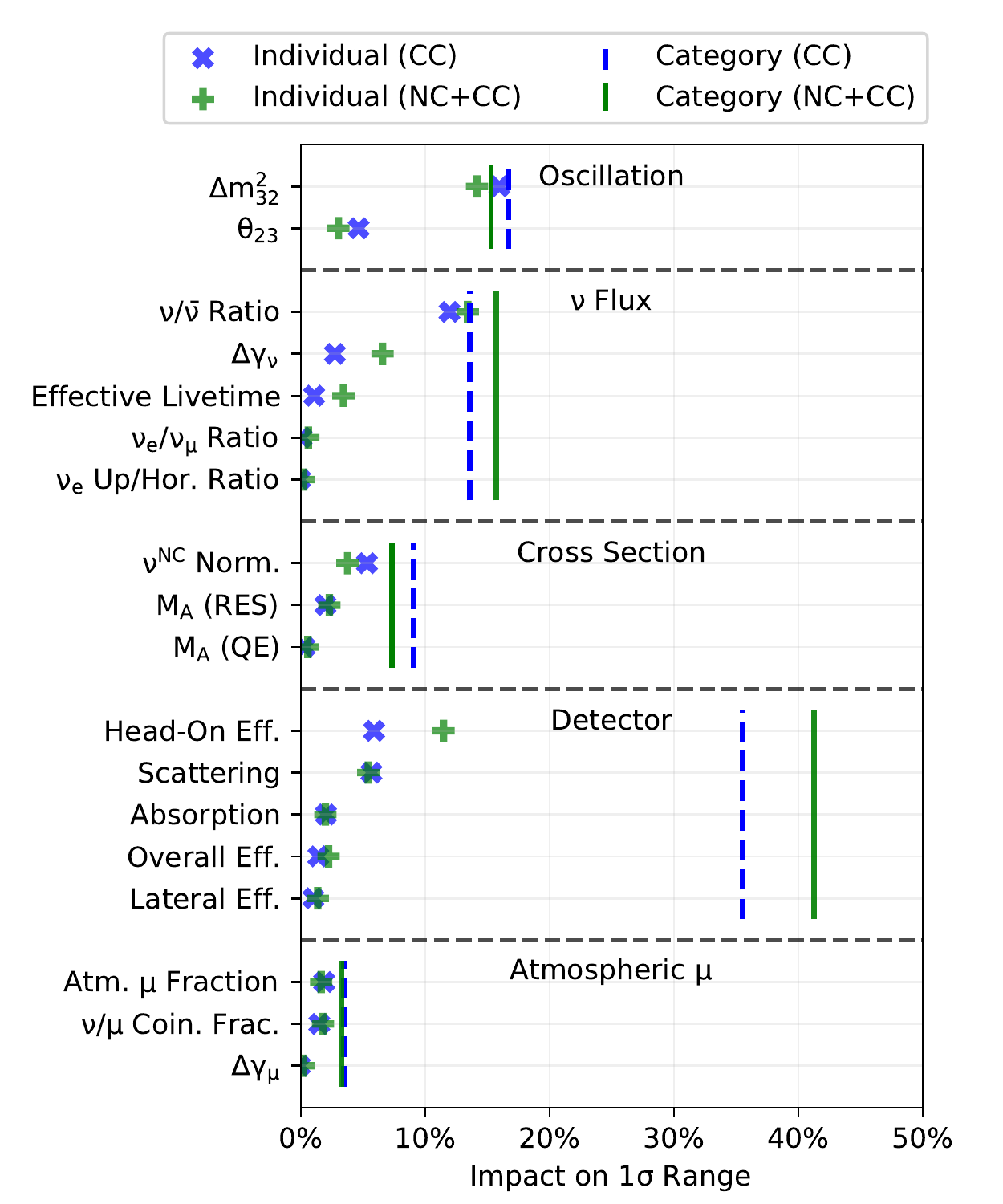}
    \end{minipage}
    \caption{Top left: The measured values for the rate of appearance with respect to the expectation from unitarity for CC+NC and CC-only results in two IceCube analyses. Bottom left: Background-subtracted $\nutau$ signal events projected in reconstructed energy, direction and event topology. Right: The relative impact from each systematic uncertainty and each group on the final 1$\sigma$ confidence interval width in IceCube's $\nutau$ appearance result. Each systematic uncertainty is fixed to the best-fit value in turn and the change in the interval is measured. Figures reproduced from~\cite{Aartsen:2019tjl}.
    }
    \label{fig:dcresults}
\end{figure}

The most recent result was obtained using a dataset corresponding to three years of detector livetime and it included a dedicated channel for cascade-like events. Such a channel is relevant because the leading oscillation effect is $\numu \rightarrow \nutau$ and, being above the $\tau$ production threshold, $\nutau$ interactions in DeepCore produce cascades. The standard oscillation picture can thus be constrained simultaneously by the disappearance of $\numu+\numubar$ and the appearance of $\nutau+\nutaubar$. Moreover, the standard oscillation picture can be tested by scaling the rate of $\nutau$ appearance with respect to the expectation from a unitary mixing matrix. Two analyses reported results from this study, with the most precise one rejecting the absence of $\nutau(\nutaubar)$CC at 2.0$\sigma$ and measuring an appearance rate of $0.57^{+0.36}_{-0.30}$ with respect to the expectation~\cite{Aartsen:2019tjl}. The results and the background subtracted signal are shown in Fig.~\ref{fig:dcresults}, left.

The search for $\nutau$ appearance is the most sophisticated analysis of DeepCore data to date. The results and the impact of the various uncertainties considered in this analysis is shown in Fig.~\ref{fig:dcresults}, right. The figure shows the expected error reduction in the appearance rate result after assuming each parameter is known perfectly. As for most low energy analyses, the imperfect knowledge of the medium and detector response are responsible for most of the error. The neutrino flux and cross section come next, with some degeneracy between them. Explicit uncertainties on the cross section are the relative scaling of neutral current to charge current interactions and the values used for the axial mass $M_A$ for resonant and quasielastic interactions. A possible deviation on the cross section as function of energy is considered by an $E^\gamma$ scaling factor, which is degenerate with uncertainties on the flux.

Various DIS cross section related uncertainties were tested but were found to either have a negligible impact on the sample or have a small impact and be very highly correlated with the neutrino flux normalization and the $E^\gamma$ correction factor. These are corrections on:
\begin{itemize}
    \item Low $Q^2$ region - As explained in Sec.~\ref{sec:lowe_xs}, GENIE uses the Bodek-Yang model to extend parton distribution functions to lower $Q^2$ regions, which DeepCore data probes. Events were re-weighted using the scheme provided by GENIE, which grants access to the higher-twist parameters and valence quark corrections. 
    \item Low $x$ region - The difference between GENIE predictions and low Bjorken-$x$ data from NuTeV~\cite{Tzanov:2005kr} was parameterized as an function of $x$ and used to reweight the simulation~\cite{Mandalia:2020lhz}.
    \item Hadron multiplicity - Visible inelasticity $y$ was calculated using an additional hadronization model designed to reproduce the averaged charged hadron multiplicity from neutrino bubble chamber data~\cite{Katori:2014fxa}. The difference was parameterized as a function of $y$ and then used to reweight the simulation.
\end{itemize}
Even though the $\nutau(\nutaubar)$-nucleon cross section has not been measured at these energies, an uncertainty on this parameter was not considered as it would have been completely degenerate with the tau neutrino interaction rate measurement. This might change in future studies with higher statistics and/or improved detectors.

\subsection{Inelasticity in the TeV scale}

\begin{figure}[h!]
\centering
\includegraphics[width=\columnwidth]{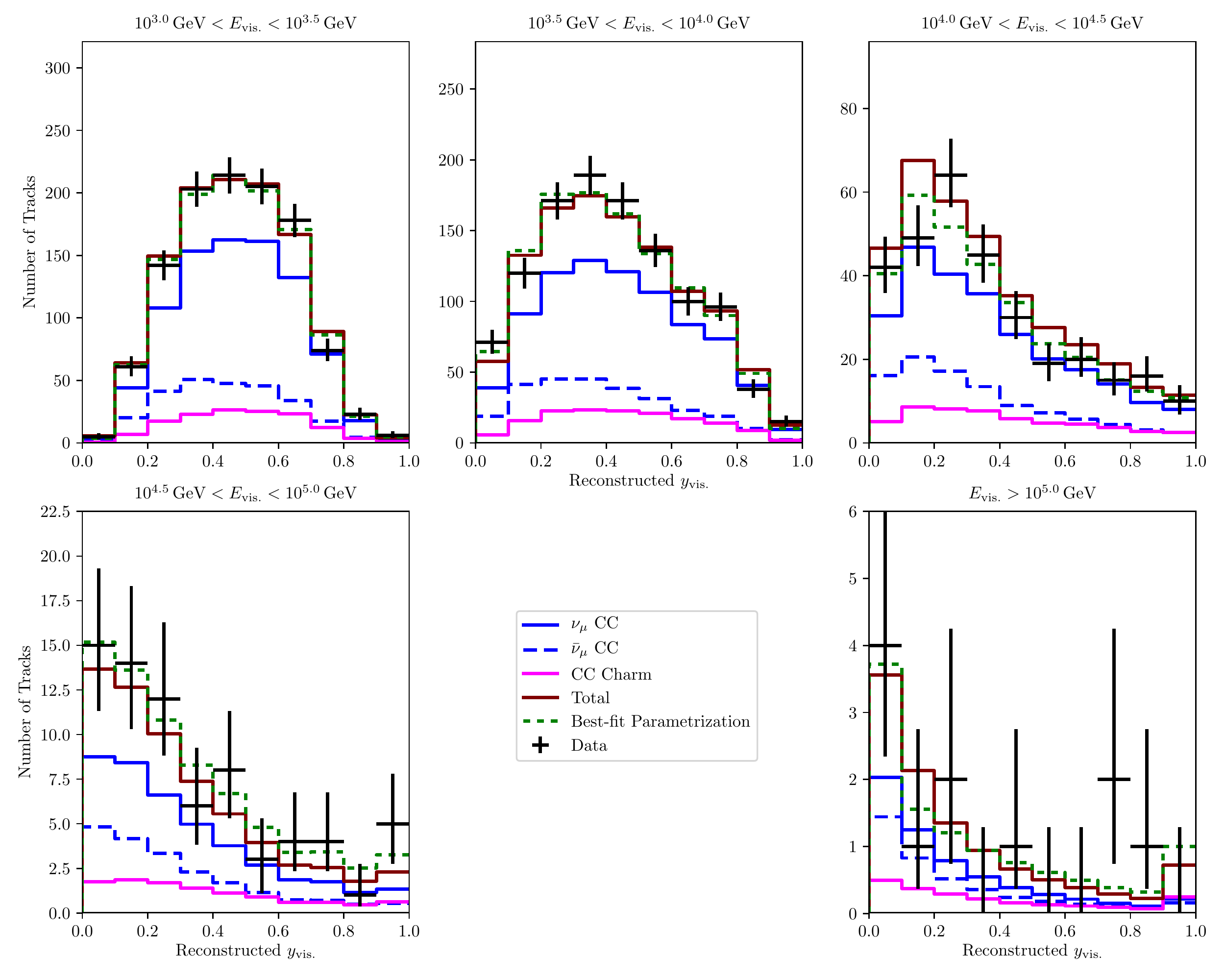}
\caption{Reconstructed visible inelasticity distributions with different visible neutrino energy bins~\cite{Aartsen:2018vez}. CSMS model predictions on $\numu$CC (blue solid), $\numubar$CC (blue dashed), CC charm (magenta), and the sum (brown) are shown. Green dashed line is from the parametrized model made from the data.}
\label{fig:inelasticity}
\end{figure}

Inelasticity ($y$) measurement is interesting because the single differential cross-section $d\sigma/dy$ is measurable by neutrino telescopes with minimum inference. Inelasticity can be used to statistically separate $\nu$ and $\nubar$ events at energy below $\sim$100 TeV, or this can be used for various PIDs~\cite{Aartsen:2018vez}. For $\numu$($\numubar$)CC interactions happening inside of the IceCube target volume, visible inelasticity $y_{vis}$ is reconstructed. Photons from the hadronic shower make a cascade energy deposit $E_{casc}$ at the interaction vertex, and the fit can find both $E_{casc}$ and the track energy deposit $E_{track}$ from the muon track. Then visible energy deposit (estimater of neutrino energy) $E_{vis}=E_{casc}+E_{track}$ and visible inelasticity $y_{vis}=E_{casc}/E_{vis}$ are reconstructed. $E_{casc}$ is lower than the true hadron shower energy deposit, this also makes both $E_{vis}$ and $y_{vis}$ to be lower than true neutrino energy and true inelasticity.

Figure~\ref{fig:inelasticity} shows the measured inelasticity distribution in different visible energy bins. Data are consistent with predictions based on the CSMS model~\cite{CooperSarkar:2011pa}. Due to the limited statistics and incomplete systematics, instead of correcting the data to obtain the $d\sigma/dy$ distribution, the data is used to fit a model to extract the true $y$ distribution. Inelasticity distributions are binned in five $E_{vis}$ regions, and they are parameterized with two parameters in a simple model ($\propto (1+\epsilon(1-y^2))y^{\lambda-1}$) motivated from the DIS double differential cross-section formula. Fits find $\lambda$ and the mean inelasticity $\left<y\right>$ where $\epsilon$ is a function of $\lambda$ and $\left<y\right>$. As shown in Fig.~\ref{fig:inelasticity}, the fit parameterization can reproduce the data well, and such a model is also consistent with CSMS model prediction.  
\begin{figure}[h!]
\centering
\includegraphics[width=0.7\columnwidth]{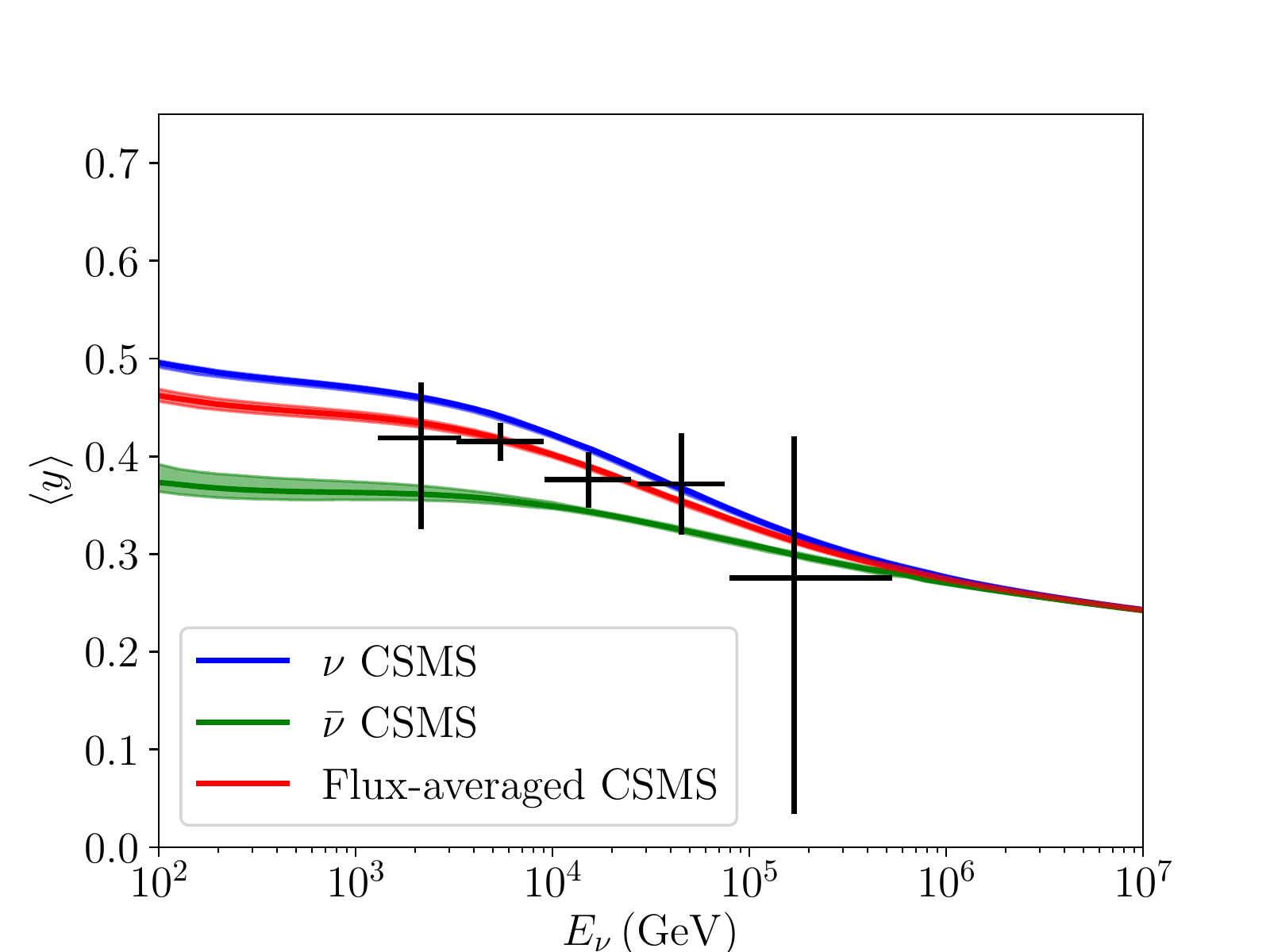}
\caption{Flux-averaged mean inelasticity distribution with function of neutrino energy. Data are compared with prediction by CSMS~\cite{CooperSarkar:2011pa}.}
\label{fig:inelasticity2}
\end{figure}

Figure~\ref{fig:inelasticity2} shows data-theory comparison of mean inelasticity as a function of neutrino energy. Here, the atmospheric neutrino flux prediction from Ref.~\cite{Honda:2006qj} is used with high-energy extrapolation ($>1$~TeV)~\cite{Aartsen:2016xlq}. As shown in Fig.~\ref{fig:inelasticity2}, the mean inelasticity is different for neutrinos and anti-neutrinos at low energy. Inelasticity is more sharply peaked at zero for anti-neutrino CC than neutrino CC interactions (Fig.~\ref{fig:inelasticity}), and after convolving flux and detector effects, this structure remains as the higher mean inelasticity for neutrino vs. anti-neutrino CC interactions. The difference decreases at higher energy. Thus this energy dependent feature can be used to break the degeneracy between atmospheric neutrino and anti-neutrino flux predictions. The best-fit value of the $\nu/\nubar$ ratio is $\nunubarRC^{\nunubarRH}_{\nunubarRL}$ where R=1 is from the simulation. Although neutrino to anti-neutrino flux ratio depends on energy and angle, this normalization-only fit shows the prediction~\cite{Honda:2006qj} is consistent with the data.

Nuclear effects are also considered for the TeV inelasticity measurement~\cite{Klein:2020nuk}. The effect is the largest at low energy ($\sim$100~GeV) and it decreases with energy. 
Nuclear effects are also largest at small $y$, but even there they are a subdominant source of systematic uncertainties. 

\subsection{Total DIS cross sections above TeV energies}

In 2013, IceCube discovered a diffuse flux of astrophysical neutrinos above \SI{60}{\tera \eV} using a sample of neutrino interactions within a contained fiducial region of the detector~\cite{Aartsen:2013jdh}. An independent confirmation was provided two years later by a separate sample of predominantly horizontal and upgoing muons from CC muon neutrino interactions outside the detector~\cite{Aartsen:2015rwa}. This presented the first opportunities to probe neutrino interactions at TeV energy scales and beyond.

\begin{table}[thb]
\centering
\begin{tabular}{l|rrrrr}
Publication & Sample & Livetime & Energy range & NBins & Flavor PID \\
\hline
Ref.~\cite{Aartsen:2017kpd} & Upgoing tracks & \SI{1}{yr} & \SIrange{6.3}{980}{\tera \eV} & 1 & $\mu$ \\
Ref.~\cite{Bustamante:2017xuy} &HESE cascades & \SI{6}{yr} & \SI{18}{\tera \eV} to \SI{2}{\peta \eV} & 4& $e$\\
Ref.~\cite{Abbasi:2020luf} & HESE ternary & \SI{7.5}{yr} & \SI{60}{\tera \eV} to \SI{10}{\peta \eV} & 4 & $e$, $\mu$, $\tau$\\
\end{tabular}
\caption{Comparison of the three cross section measurements performed with IceCube data. All analyses fixed $\sigma^{\rm CC}/\sigma^{\rm NC}$ and $\sigma_\nu/\sigma_{\bar{\nu}}$ ratios based on the Standard Model predictions. In addition, $y^{\rm NC}=0.25$ was assumed in Ref.~\cite{Bustamante:2017xuy}. Note that the $e$ symbol in the flavor PID column includes cascades from NC interaction channels as those are indistinguishable from CC interactions of electron (anti)neutrinos.}
\label{tab:xscomp}
\end{table}

\begin{figure*}[h!]
\centering
    \subfloat{
        \includegraphics[width=0.45\columnwidth]{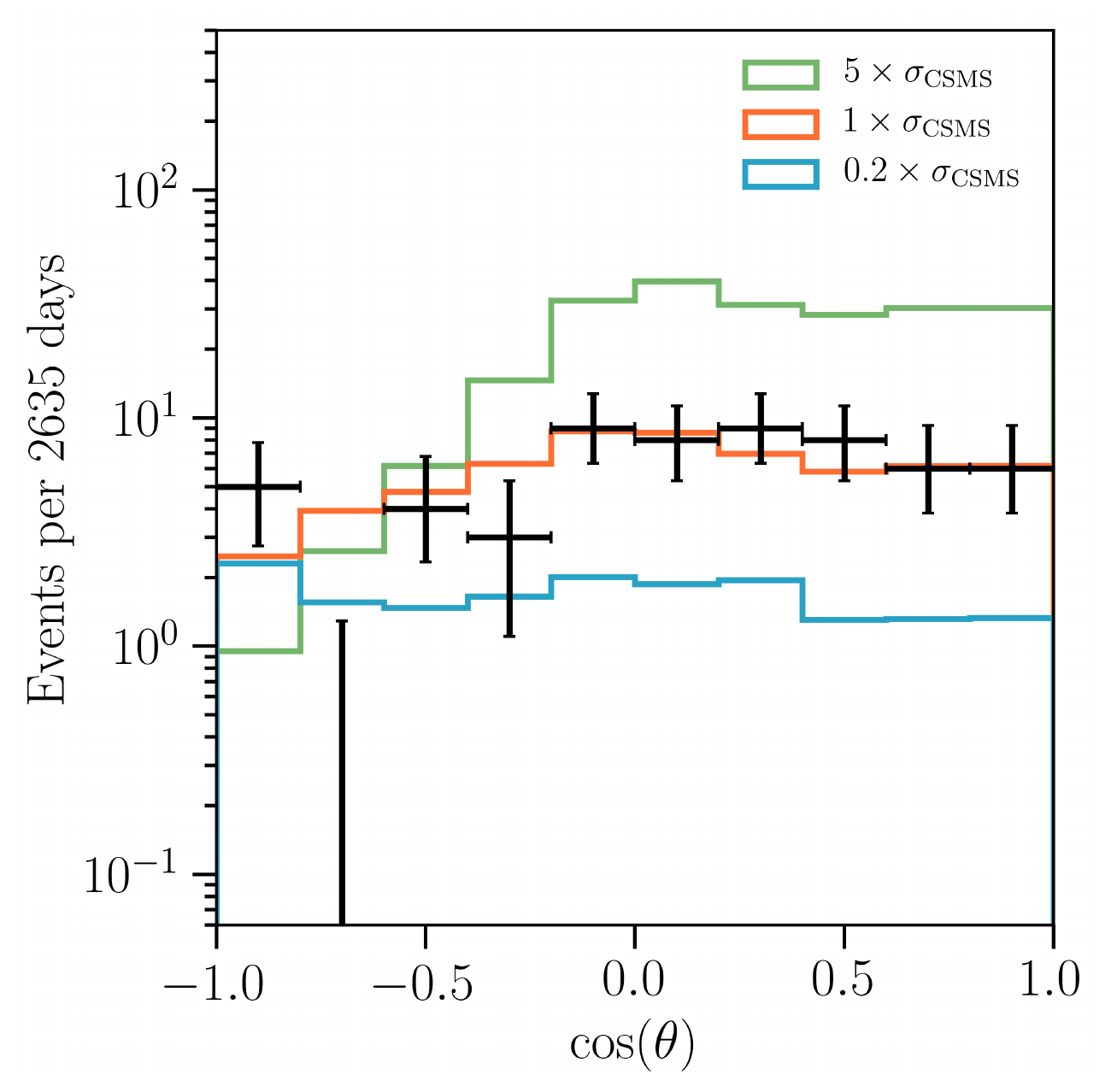}
    }
    \subfloat{
        \includegraphics[width=0.45\columnwidth]{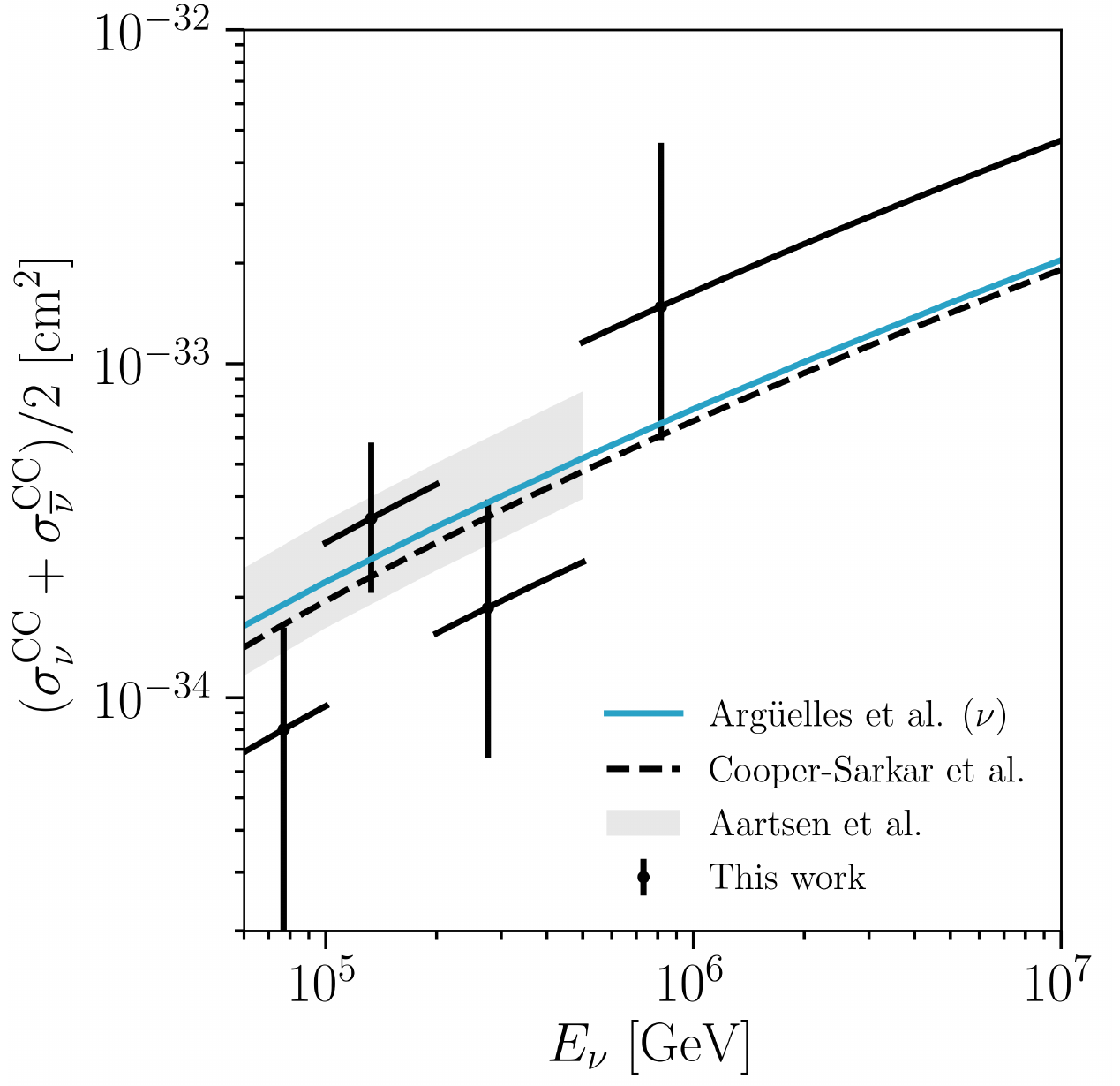}
    }

\caption{Figure reproduced from Ref.~\cite{Abbasi:2020luf}. Left panel: Zenith distribution of observed events (black error bars) in the HESE selection above \SI{60}{\tera \eV} compared to MC expectation rates from three different settings of $\sigma_{\rm CSMS}$ (blue, orange, and green lines). The effect of a scaled DIS cross section is linear in the southern sky ($\cos \theta > 0$) while also changing the arrival flux from the northern sky ($\cos \theta < 0$). Right panel: The black error bars show the frequentist result obtained from fitting the cross section scaling parameters in four different energy bins to the data. The result from Ref.~\cite{Aartsen:2017kpd} is shown as the shaded gray region. Two predictions from Ref.~\cite{CooperSarkar:2011pa} and Ref.~\cite{Arguelles:2015wba} are shown as the dashed-black line and solid-blue line, respectively.}
\label{fig:hese}
\end{figure*}

IceCube measures the total neutrino cross section as a function of energy by assuming a single-power-law flux of astrophysical neutrinos. Table~\ref{tab:xscomp} summarizes the three published cross section measurements using IceCube data. As the dominant interaction channel at these energies is DIS, the CSMS model~\cite{CooperSarkar:2011pa} is taken as a baseline model. By allowing the CSMS cross section to scale up or down, a modification in the expected event rate is observed in MC as a function of the reconstructed energy and zenith angle. The zenith angle is defined in IceCube detector coordinates such that $\cos(\theta) = 1(-1)$ corresponds to a down(up)-going neutrino. Figure~\ref{fig:hese} is from the high-energy starting event (HESE) selection sample where  
the left panel shows the expected rate for three different scalings of the CSMS cross section, $\sigma_{\rm CSMS}$ marginalizing over the reconstructed energy~\cite{Abbasi:2020luf}. The data shown as black error bars is described in detail in~\cite{Abbasi:2020jmh} and covers the full zenith range. Two important features are noticeable immediately from the figure. First, in the Southern sky ($\cos(\theta) > 0$) the effect of a scaled cross section is approximately linear. Second, in the Northern sky ($\cos(\theta) < 0$) the effect of scaling the cross section affects the shape of the expectation rate via the Earth attenuation effect such that a $5 \times \sigma_{\rm CSMS}$ cross section falls off much more steeply as a function of the chord length in the Earth. The cross section can thus be measured by finding the scaling that best fits the data.

\subsubsection{Measurement using throughgoing tracks}

\begin{figure}[h!]
\centering
\includegraphics[width=0.70\columnwidth]{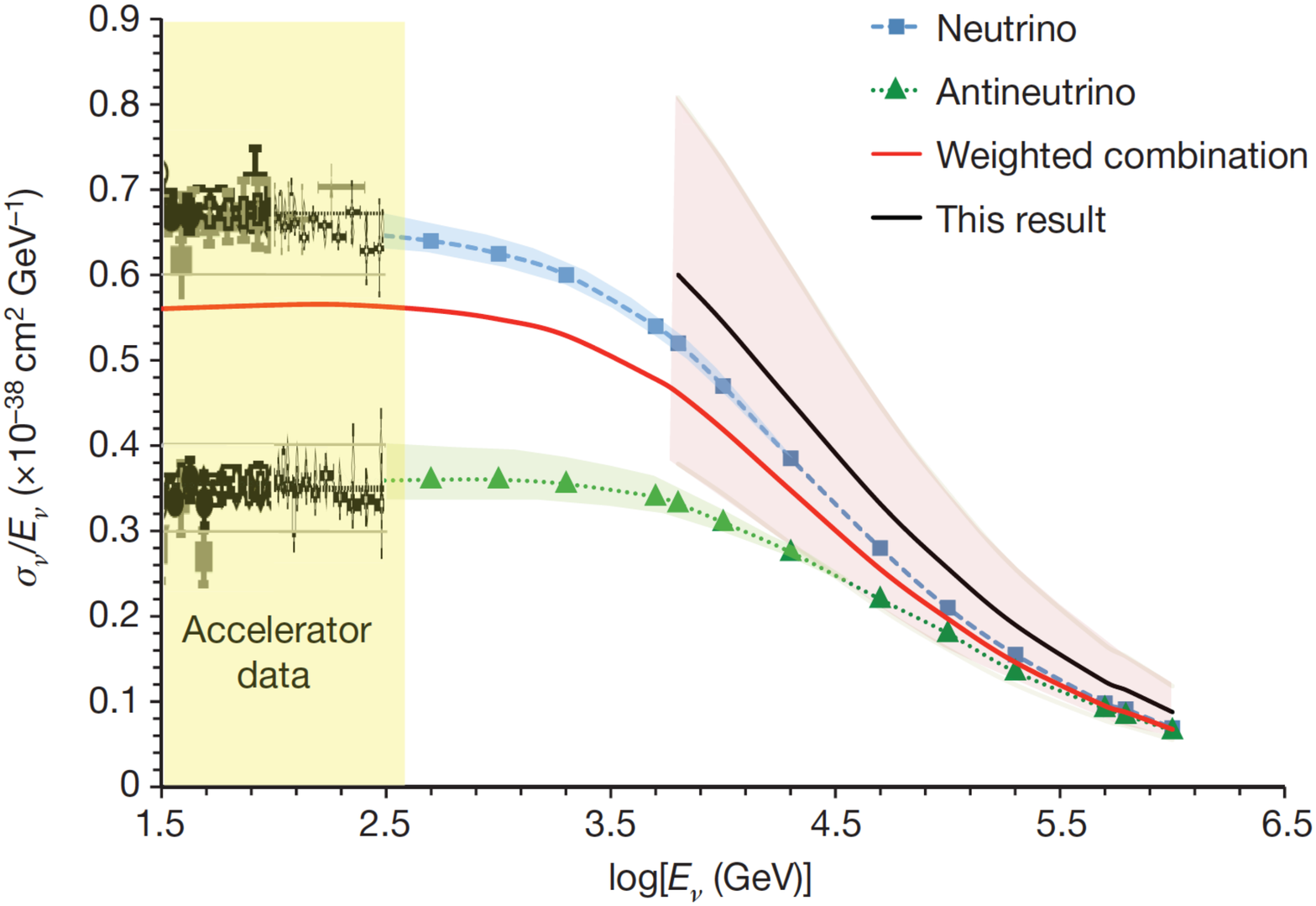}
\caption{Total muon (anti)neutrino cross section measured by IceCube using a sample of predominantly upgoing and horizontal muon neutrinos~\cite{Aartsen:2017kpd}. The result of $\tgmxs$ times $\sigma_\mathrm{CSMS}$ is applied to a weighted combination (red line) of $\sigma_\nu$ and $\sigma_{\bar{\nu}}$ and shown as the black line and red shaded region.}
\label{fig:tgmxs}
\end{figure}

In Ref.~\cite{Aartsen:2017kpd}, an overall cross section measurement was performed with a 79-string configuration from 2010-2011 based on a subset of the events in Ref.~\cite{Aartsen:2015rwa}. In this analysis, $\Nabs$ up-going high-energy muons are detected with less than 0.1\% background from the 79-string configuration in bins of reconstructed zenith angle and energy. Although the angle can be measured better than $\absangres$, throughgoing muons are not entirely contained in the detector and the neutrino energy is estimated from $dE/dx$ and MC to within roughly a factor of $\absengres$. Assuming a fixed CC-to-NC cross section ratio, a binned forward-folding fit finds a cross section of $\tgmxs$ times that of the Standard Model in the energy range \SIrange{6.3}{980}{\tera \eV}. The result, shown in Fig.~\ref{fig:tgmxs}, exhibits a deviation of the DIS cross-section from a linear increase with energy at around 3~TeV, due to the suppression of the finite gauge boson mass in the propagator ($\sim(Q^2+M_W^2)^{-1}$). Although this is known Standard-Model effect, it is the first time to be confirmed in neutrino interactions. 

While the sample is dominated by muons induced from muon neutrino CC interactions in the region surrounding the instrumented region, NC interactions affect the event rate by cascading neutrinos down to a lower energy thus changing the arrival flux at IceCube. The astrophysical flux is centered on the result from~\cite{Aartsen:2015knd}, which combines six previously studied samples in a maximum-likelihood fit of the diffuse astrophysical neutrino flux. Systematic uncertainties on the astrophysical and atmospheric neutrino flux, global energy scale, ice modeling, density of the Earth, atmospheric pressure, and angular acceptance of photosensors were taken into account~\cite{Aartsen:2017kpd}. However, the method by which they were evaluated differ. First, eight parameters describing the conventional, prompt, and astrophysical normalizations (3), the conventional and astrophysical spectral index (2), kaon-to-pion ratio (1), muon neutrino-to-antineutrino ratio (1), and global energy scale uncertainty (1) were included as nuisance parameters directly in the fit. A total uncertainty of $1.30^{+0.30}_{-0.26}$ was obtained. Next, the statistical-only uncertainty was factored out by keeping all nuisance parameters fixed at their best-fit values. Finally, the remaining systematics (ice modeling, Earth density, atmospheric pressure, and angular acceptance) were evaluated individually and added in quadrature on top of the systematic uncertainty from the fit result to obtain the total uncertainty due to systematics.

\subsubsection{Measurement using contained cascades}

Following this result, which relies predominantly on tracks, a measurement of the neutrino DIS cross section was performed using a sample of contained cascades~\cite{Bustamante:2017xuy} in the HESE selection with six years of public IceCube data~\cite{Aartsen:2013bka,Kopper:2017Df,Kopper:2016DR}. Since cascades have better energy resolution the idea was to probe the DIS cross section as a function of neutrino energy in the range \SIrange{18}{2000}{\tera \eV}. The analysis approximated the arrival flux at IceCube by assuming an exponential attenuation of the neutrino flux as a function of $E_\nu$ and zenith angle. Four energy bins were constructed and the astrophysical flux normalization and spectral index were allowed to vary in each energy bin while assuming equal flavor breakdowns. The atmospheric conventional flux model assumed is from Ref.~\cite{Honda:2006qj}, while the prompt flux was fixed to zero. Since a full MC was not available for this analysis, the detector event rate is computed numerically and a \SI{10}{\percent} (\SI{15}{\degree}) energy (angular) uncertainty was assumed to simulate the detector response. Three additional simplifying assumptions were made: first the inelasticity is set to 0.25 for NC interactions, the CC-to-NC cross section ratio is fixed to 3, and third the neutrino to antineutrino cross section ratio is fixed in each bin.

\subsubsection{Measurement using contained events}

A DIS cross section measurement obtained using the entire high-energy starting event (HESE) selection with 7.5 years of data~\cite{Abbasi:2020jmh,Abbasi:2020luf} was the first to include all three flavor proxies in the analysis~\cite{Usner:2018cel}. The contained selection consisted of 60 events above \SI{60}{\tera \eV}, and the cross section was measured in four energy bins with edges at 60~TeV, 100~TeV, 200~TeV, 500~TeV, and 10~PeV. Four scaling parameters, $\bm{x}=(x_0, x_1, x_2, x_3)$, linearly scale the CSMS cross section in these four energy bins respectively. Each event in the MC simulation can then be reweighted by $x_i \Phi(E_\nu, \theta_\nu, \bm{x})/\Phi(E_\nu, \theta_\nu, \bm{1})$, where $\Phi$ is the arrival flux as calculated by \nusquids{}~\cite{Delgado:2014kpa}, $E_\nu$ is the true neutrino energy, $\theta_\nu$ the true neutrino zenith angle, and $x_i$ the cross section scaling factor at $E_\nu$. Similar to the other analyses, a fixed CC-NC cross section ratio is assumed. Additionally, the astrophysical neutrino to antineutrino flux ratio is fixed to unity, though its effects were studied and found to be negligible.

A forward-folding fit is then performed in the reconstructed energy vs zenith distribution for tracks and cascades, and in the reconstructed energy vs cascade length separation distribution for double cascades. All four cross section parameters are simultaneously fitted in both the frequentist and Bayesian statistical paradigms. Systematic uncertainties on the astrophysical and atmospheric neutrino flux were included, with the largest contribution due to the astrophysical spectrum. Detector systematics were studied and found to be negligible. The result is shown as the black error bars in Fig.~\ref{fig:hese} with the result from Ref.~\cite{Aartsen:2017kpd} shown as the shaded grey band. The CSMS prediction is shown as the black dashed line, while an alternative calculation from Ref.~\cite{Arguelles:2015wba} is shown as the blue solid line for $\sigma_\nu^{\rm CC}$. The error bar is drawn to follow the curve of the CSMS prediction within each energy range. 

\subsection{On-shell $W$-boson production}

Much of the preceding discussion has focused on $\nu N$ interactions, either via DIS or resonance and quasielastic processes depending on the neutrino energy. The on-shell production of a $W$-boson is another interaction channel with non-negligible contributions to neutrino interactions in large scale neutrino telescopes. For an electron antineutrino scattering off an atomic electron, the resonance peaks at $E_\nu=\SI{6.3}{\peta \eV}$~\cite{Glashow:1960zz,Loewy:2014zva}. It was first predicted by Sheldon Glashow in 1959~\cite{Glashow:1960zz} and later proposed as a method to search for $W$-bosons with large-volume, underground Cherenkov detectors (neutrino telescopes)~\cite{Berezinsky:1977sf}. A particle shower detected by IceCube provides the first detection of this process at the $2.3\sigma$ level~\cite{IceCube:2021rpz}.

\begin{figure}[h!]
\begin{center}
    \includegraphics[width=\columnwidth]{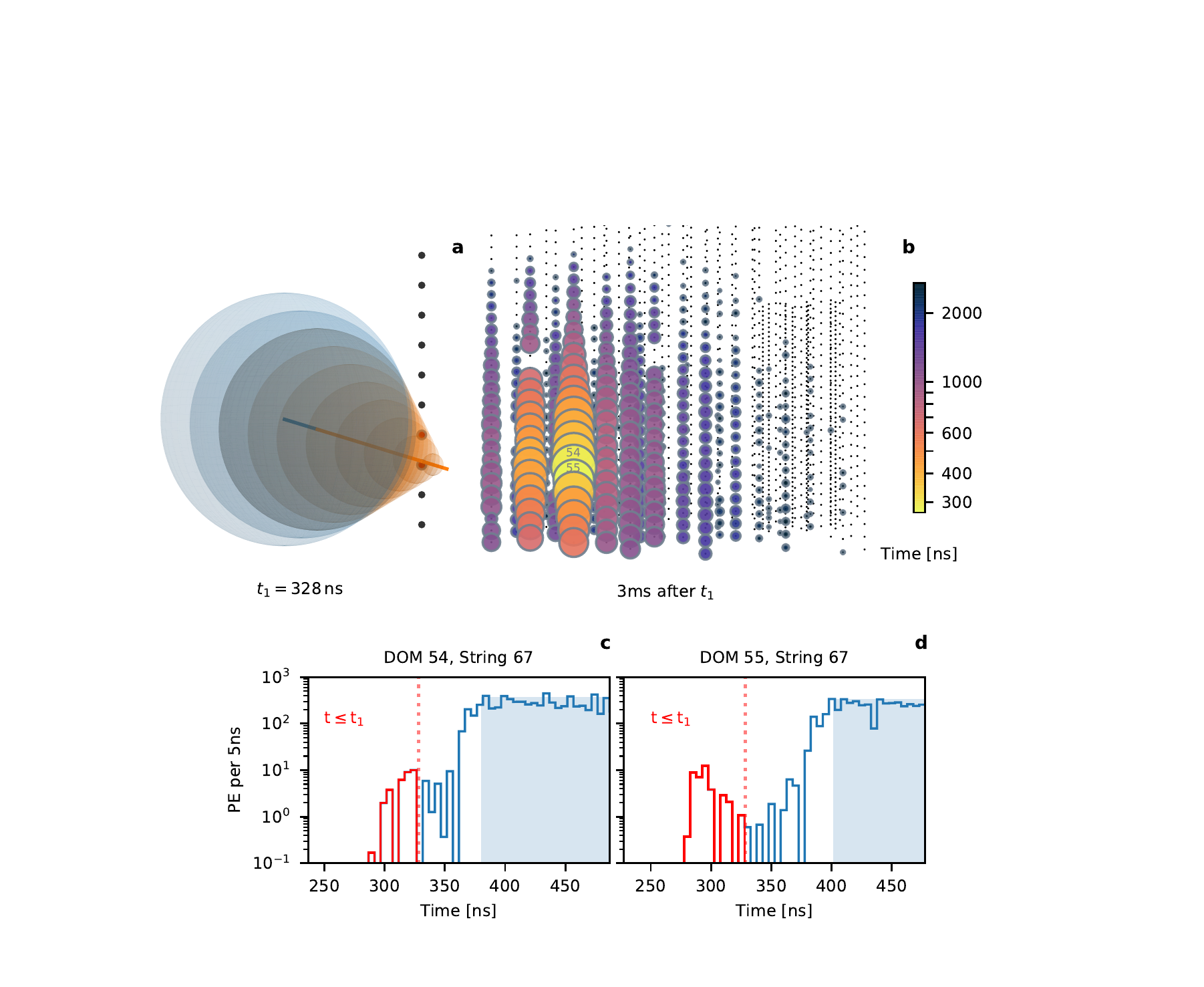}
    \caption{Adapted from Ref.~\cite{IceCube:2021rpz}. The top-left panel (a) shows the inferred direction of the $\bar{\nu}_e$-induced shower (blue line) and muon (orange line), along with their Cherenkov wavefronts shown as shaded spheres in the same color scheme. The top-right panel (b) shows the final development of the event over a large portion of the detector, with the size of the circles corresponding to the total charge observed by that DOM and the color indicative of the first-photon arrival time. The bottom two panels (c and d) show the extracted photoelectron distributions on the two closest DOMs, with the early pulse distribution shown in red. The blue shaded region indicates when the PMT was saturated.}
    \label{fig:grevent}
\end{center}
\end{figure}

Experiments are most sensitive to the $W^-$ or Glashow resonance  when the $W^-$ decays hadronically. Leptonic decays miss energy carried away by the outgoing neutrino, and the outgoing lepton energy is smeared over a broader range. Thus, cascades offer the best chance of isolating the $W$-resonance. The contained searches in IceCube did not yield any events with energy above \SI{2}{\peta \eV}~\cite{Abbasi:2020jmh}. A partially-contained selection increases the detection volume by roughly a factor of two compared to the contained search and upon unblinding one cascade was observed with reconstructed visible energy above \SI{4}{\peta \eV}~\cite{IceCube:2021rpz}. The event vertex was reconstructed approximately \SI{80}{\m} outside the detector. Uniquely, three DOMs on the nearest string detected pulses that arrived earlier than possible assuming the speed light travels in ice. These early pulses are shown as red histograms in the two lower panels of Fig.~\ref{fig:grevent}, where the red dotted line depicts the earliest time light from the shower could have arrived on the two DOMs. The upper panels of Fig.~\ref{fig:grevent} show event displays at two time slices: \SI{328}{\nano \s} after the inferred interaction time (top right) and \SI{3}{\milli \s} following the inferred interaction time. The top left panel indicates that Cherenkov radiation from a muon (orange cone and orange line) can outrun that of the shower (blue shaded region), and thus produce the early pulses. A two-step reconstruction of the shower and muon properties rejects the atmospheric background hypothesis at the $5\sigma$ level. Due to large systematic uncertainties in hadronic interaction models, the neutrino DIS background rejection was evaluated based on the reconstructed energy of the shower alone to be $2.3\sigma$.

It is also now well known that the s-channel process described by Glashow is not the only means to obtain $W$-bosons. Neutrino interactions off the photon field of the nucleus occur for all three neutrino flavors and can also yield on-shell $W$s~\cite{Seckel:1997kk,Alikhanov:2015kla,Zhou:2019vxt}. This has been referred to as the ``hidden Glashow'' process, though it does not result in a Breit-Wigner-like peak and is also referred to simply as $W$-boson production. For leptonic decays of the $W$, the final outgoing particles can consist of two leptons of opposite sign and is referred to as ``trident'' production~\cite{Zhou:2019vxt,Ge:2017poy}. Importantly, tridents have also been discussed in the context of accelerator neutrino experiments~\cite{Altmannshofer:2014pba,Magill:2016hgc,Ballett:2018uuc,Altmannshofer:2019zhy}. While these Standard Model signatures have not been observed experimentally, prospects for their detection in IceCube and future large-volume neutrino telescopes appear promising~\cite{Beacom:2019pzs}.

\section{Future prospects}
\label{sec:future}

\subsection{IceCube-Upgrade}

The IceCube Upgrade is a funded extension to the IceCube Neutrino Observatory that features about 700 new photodetectors and new calibration devices to be deployed within the DeepCore volume \cite{Ishihara:2019aao}. The additional instrumentation will push the detector energy threshold down to about 1~GeV and at higher energies it will provide more information per event. Simulation studies performed with a similar level of sophistication as current DeepCore results suggest the Upgrade can deliver a 10\% precision measurement of the $\nutau$ appearance rate with a single year of data \cite{Stuttard:2020zsj}. This appearance rate will be a significant contribution to the unitarity test of the neutrino mixing matrix. Moreover, the gain in statistics and resolution will also make it possible to turn this appearance rate measurement into a total cross section measurement $\sigma(E_\nu)$ of tau neutrinos. Deployment is expected to happen in 2022-2023.

\subsection{IceCube-Gen2}

IceCube-Gen2 is a future project with three parts. A highly-instrumented low-energy array (IceCube Upgrade) will be used to study the neutrino mass ordering through neutrino oscillations~\cite{IceCube:2018ikn,IceCube:2019dyb}. A sparsely instrumented high-energy extension array and a surface radio array will measure high-energy neutrinos. The high-energy extension detector is envisioned to increase the instrumented in-ice volume by approximately a factor of eight. This will be accomplished by drilling over 100 additional strings, spaced \SIrange{200}{300}{\m} apart, with upgraded optical modules and calibration devices~\cite{Aartsen:2020fgd}. The larger volume will increase the IceCube effective area by approximately a factor of five, increasing the number of neutrino interactions by the same proportion and making it possible to probe neutrino cross sections at energies of \SI{10}{\peta \eV} and above. Uncertainties on existing cross section and inelasticity measurements will be significantly reduced by IceCube-Gen2. A planned surface radio array capable of detection Askaryan radiation~\cite{Saltzberg:2000bk,Prohira:2019glh} from particle showers will extend the energy reach to the EeV scale, and it should become possible to effectively probe new physics with neutrino cross sections~\cite{Klein:2019nbu}. Such models predict an enhancement of the neutrino interaction rate over CC+NC DIS in the context of large extra dimensions~\cite{Jain:2000pu,AlvarezMuniz:2002ga}, leptoquarks~\cite{Romero:2009vu} and sphalerons~\cite{Ellis:2016dgb}.

\subsection{Other neutrino telescopes}

Although this review focuses on neutrino cross-section analyses in IceCube, other neutrino telescopes also can contribute this subject significantly. First, neutrino detectors based on target materials in tanks such as Super-Kamiokande~\cite{Desai:2007ra}, and future DUNE~\cite{Abi:2020wmh} and Hyper-Kamiokande~\cite{Abe:2018uyc} can measure neutrino interactions up to several TeV~\cite{Desai:2007ra}. The main drawback is the small volume compared with experiments without tanks, and the event rate falls off quickly due to low conventional atmospheric neutrino flux at higher energy ($\sim E^{-3.7}$). Nevertheless, these experiments achieve better angle and energy resolutions, and they are superior to IceCube in measurements of low energy neutrino interactions  precisely including NC quasi-elastic scattering cross-section on oxygen~\cite{Wan:2019xnl} and $\nutau$CC cross-section~\cite{Li:2017dbe}. 

Second, future Cherenkov arrays in natural bodies of water, such as KM3NeT~\cite{Adrian-Martinez:2016fdl}, Baikal-GVD~\cite{Safronov:2020dtw}, and  P-ONE~\cite{Agostini:2020aar} can measure high-energy neutrino interactions with similar statistics as IceCube. KM3NeT has a similar optical sensor array as IceCube, where 3 separated building blocks host $\mDOMString$ strings each, and each string has $\mDOMperString$ optical modules. The main difference from others is each optical module consists of $\mDOMNPMT$ 3-inch PMTs with $4\pi$ coverage. This approach is expected to measure down-going events more precisely where high-energy astrophysical neutrino events are mainly down-going or horizontal events. On the other hand, the P-ONE design has multiple clusters of strings, instead of uniformly distributed strings, to maximize sensitivity to horizontal track events. Due to relatively simpler photon propagation in water than ice, in general water based experiments also have better angular resolution. The low energy array of KM3NeT, ORCA, can measure GeV range neutrino cross-section including the $\nutau$CC cross section~\cite{Adrian-Martinez:2016fdl}. 

Third, radio telescope experiments and air shower experiments can extend measurement to the EeV ($10^{18}$~eV) region~\cite{Aartsen:2020fgd,stephanie_wissel_2020_4123869}. Radio telescopes measure the Askaryan effect~\cite{Saltzberg:2000bk,Prohira:2019glh}, the coherent radio wave emission from high-energy interactions, and experiments can cover much larger volumes than Cherenkov photon optical sensor arrays. ANITA demonstrated its radiowave measurements from cosmic shower as well as up-going neutrino candidate events~\cite{Gorham:2018ydl}. 
On the other hand, air shower experiments target radio or fluorescence emission mainly from high-energy tau leptons produced by high-energy tau neutrinos skimming the limb of the Earth. These experiments can reconstruct primary lepton kinematics, and future high statistics data allow to measure neutrino cross-sections in the EeV region~\cite{Denton:2020jft}.

\vspace{1cm}

Neutrino telescopes observe neutrinos at energies well beyond what can be achieved in a laboratory and thus are a unique tool to probe neutrino interactions. The conditions of the experiments, deployed in a natural medium and relying on naturally occurring fluxes, make these measurements challenging. However, IceCube has demonstrated that thanks to the statistics it collects and by using advanced analysis methods it is possible to probe the Standard Model and search for new physics in this area. Moreover, the multiple neutrino telescopes in construction and being planned will further extend the energy range and improve the precision of these studies over the next couple of decades. 

\section*{Acknowledgements}
We thank the IceCube collaboration for the careful reading of this manuscript. The authors gratefully acknowledge the support
from National Science Foundation (USA), Arthur B. McDonald Canadian Astroparticle Physics Research Institute (Canada), Science and Technology Facilities Council (UK).

\bibliographystyle{epj}
\bibliography{IceCube}

\end{document}